\documentclass{jpp}
\usepackage{amsfonts}   
\usepackage{amsmath}    
\usepackage{amssymb}    %
\usepackage{array}      
\usepackage{bm}         
\usepackage{booktabs}   
\usepackage{caption}    
\usepackage{color}      
\usepackage{graphicx}   
\usepackage[            %
    colorlinks,         %
    urlcolor = blue,    %
    citecolor = blue,   %
    linkcolor = blue    %
    ]{hyperref}         
\usepackage{lscape}     
\usepackage{mathtools}  
\usepackage{natbib}     
\usepackage{psfrag}     
\usepackage{pstool}     
\usepackage{rotating}   
\usepackage{tensor}     
\usepackage{ulem}       
\usepackage{tensor}

\providecommand{\tnsr}[5]{\tensor*[^{#2}_{#3}]{#1}{^{#4}_{#5}}}

\providecommand{\tRr}[3]{\tnsr{#1}{}{}{#2}{#3}}

\providecommand{\tR}[2]{\tnsr{#1}{}{}{#2}{}}
\providecommand{\tr}[2]{\tnsr{#1}{}{}{}{#2}}

\allowdisplaybreaks

\providecommand{\mT}[3]{\tensor{#1}{_{#2#3}}}   

\DeclareMathOperator{\sgn}{sgn}
\DeclarePairedDelimiter\abs{\lvert}{\rvert}%
\DeclarePairedDelimiter\norm{\lVert}{\rVert}%
\makeatletter
\let\oldabs\abs
\def\abs{\@ifstar{\oldabs}{\oldabs*}}
\let\oldnorm\norm
\def\norm{\@ifstar{\oldnorm}{\oldnorm*}}
\makeatother

\ifCUPmtlplainloaded \else
  \checkfont{eurm10}
  \iffontfound
    \IfFileExists{upmath.sty}
      {\typeout{^^JFound AMS Euler Roman fonts on the system,
                   using the 'upmath' package.^^J}%
       \usepackage{upmath}}
      {\typeout{^^JFound AMS Euler Roman fonts on the system, but you
                   dont seem to have the}%
       \typeout{'upmath' package installed. JPP.cls can take advantage
                 of these fonts, if you use 'upmath' package.^^J}%
      }
  \else
  \fi
\fi

\ifCUPmtlplainloaded \else
  \checkfont{msam10}
  \iffontfound
    \IfFileExists{amssymb.sty}
      {\typeout{^^JFound AMS Symbol fonts on the system, using the
                'amssymb' package.^^J}%
       \usepackage{amssymb}%
       \let\le=\leqslant  \let\leq=\leqslant
         
      }{}
  \fi
\fi

\DeclareRobustCommand{\vect}[1]{\bm{#1}}
\ifCUPmtlplainloaded \else
  \IfFileExists{amsbsy.sty}
    {\typeout{^^JFound the 'amsbsy' package on the system, using it.^^J}%
     \usepackage{amsbsy}}
    {}
\fi

\newsavebox{\astrutbox}
\sbox{\astrutbox}{\rule[-5pt]{0pt}{20pt}}

\providecommand{\aV}[1]{\left\langle{#1}\right\rangle}

\def\<{\aV{}
\def\>{}}

\def\half{{\textstyle{1\over2}}}

\def\rmd{d}
\def\d{\rmd}
\def\rmi{i}

\def\be{\begin{equation}}
\def\ee{\end{equation}}
\def\bea{\begin{eqnarray}}
\def\eea{\end{eqnarray}}
\def\nn{\nonumber}

\newfont{\myfont}{cmmib10}

\newcommand{\bzeta}{\hbox{\myfont \symbol{16} }}

\title[Response of a Relativistically Streaming Pulsar Plasma]{Response of a Relativistically Streaming Pulsar Plasma}

\author[M. Z. Rafat, D. B. Melrose, and V. M. Demcsak]%
{M.\ns Z.\ns  R\ls A\ls F\ls A\ls T$^{1}$, D.\ns B.\ns M\ls E\ls L\ls R\ls O\ls S\ls E$^1$%
  \thanks{Email address for correspondence: donald.melrose@sydney.edu.au},
\and V.\ns M.\ns  D\ls E\ls M\ls C\ls S\ls A\ls K$^{1}$%
\ns  }
\affiliation{$^1$SIfA, School of Physics, The University of Sydney, NSW 2006, Australia
}

\pubyear{?}
\volume{?}
\pagerange{?}
\date{?; revised ?; accepted ?. - To be entered by editorial office}
\begin{document}

\maketitle

\begin{abstract}
The response tensor  is derived for a relativistically streaming, strongly magnetized, one-dimensional J\"uttner distribution of electrons and positrons, referred to as a pulsar plasma. This is used to produce a general treatment of wave dispersion in a pulsar plasma. Specifically, relativistic streaming, the spread in Lorentz factors in a pulsar rest frame, and cyclotron resonances are taken into account. Approximations to the response tensor are derived by making approximations to  relativistic plasma dispersion functions appearing in the general form of the response tensor. The cold-plasma limit, the highly relativistic limit, and limits related to cyclotron resonances are considered. The theory developed in this paper has applications to generalised Faraday rotation in pulsars and magnetars. 
\end{abstract}


\section{Introduction}

It is widely assumed that the plasma in a pulsar magnetosphere is created in pair cascades \citep{HA01,AE02,ML10,TA13}. Due to the extremely strong magnetic field, the electrons and positrons quickly radiate away all their perpendicular energy, so that they are in one-dimensional (1D) motion along the magnetic field lines. These properties favor a model for a 1D J\"uttner (relativistic Maxwellian) distribution for the pairs in a pulsar plasma \citep{HA01,AE02,ML10,TA13}. Such a 1D distribution is of the form $\propto\exp(-\rho\gamma)$ in its rest frame where $\rho$ is the inverse temperature (in units of the rest energy of the electron),  $\gamma=(1-\beta^2)^{-1/2}$ is the Lorentz factor, and $\beta$ is the speed (in units of speed of light $c$).  \cite{AE02} also suggested a value of $ \rho $ between about 0.1 and 1, with the distribution streaming with Lorentz factor $ \gamma_{\rm s}$ between about $10^2$ and $10^3 $. We refer to such a plasma, that is, a relativistically streaming, 1D  J\"uttner distribution of electron-positron pairs, as a ``pulsar plasma''. In most discussions of the response of a pulsar plasma, the wave frequency $\omega$ is assumed much smaller than the electron cyclotron frequency $\Omega_e=eB/m$, where $B$ is the magnetic field strength, $e$ is the elementary charge, and $m$ is the mass of the electron. This low-frequency limit is relevant to models in which the pulsar radio emission is generated at relatively low heights in the magnetosphere, such that the contribution of the cyclotron resonances to the plasma dispersion is negligible. In this article we present results for the plasma response, in terms of the components of the dielectric tensor $K_{ij}(\omega,\bm{k})$, where $ \bm{k} $ is the wave vector, for a pulsar plasma including the cyclotron resonances, allowing us to discuss propagation of the radio waves through the outer regions of the magnetosphere.

We follow \cite{RMM1,RMM2} in assuming that the streaming motion is included by assuming a 1D J\"uttner distribution in the plasma rest frame and Lorentz transforming to the frame in which it is streaming. We assume that the distributions of electrons and positrons, labeled as species $\epsilon=\mp$, respectively, have number densities $\tR{n}{\epsilon}$, which may be different, but both have the same streaming speed $\beta_{\rm s}$, later relaxing this assumption to allow different streaming speeds $ \tRr{\beta}{\epsilon}{\rm s} $. A difference in the number densities implies a nonzero charge density, $\eta$, and a difference in streaming speeds implies a nonzero current density, $\bm{J}$. Both $\eta$ and $\bm{J}$ are nonzero in a pulsar magnetosphere, and both contribute to the ellipticity of the polarization of the wave modes. We denote the pulsar frame as $\mathcal{K}'$, and the rest frame of species $\epsilon$ as $\tr{\mathcal{K}}{\epsilon}$, replaced by the rest frame $\mathcal{K}$ of the plasma when the streaming speeds are the same.

Dispersion in a collisionless plasma is associated with resonances. The gyroresonant frequencies satisfy $\omega-s\Omega_e/\gamma-k_\parallel v_\parallel=0$, or $z-sy/\gamma-\beta=0$, with $ \beta = v_\parallel / c $, $ z = \omega / c k_\parallel $ and $ y = \Omega_e / c k_\parallel $, where $s$ is the gyroresonant number, and subscripts $\parallel,\perp$ denote components of wave vector $\bm{k}$ and particle velocity $\bm{v}$ parallel and perpendicular to the magnetic field, respectively. In the 1D case ($v_\perp=0$), the only allowed gyroresonances are the Cerenkov resonance and the normal and anomalous cyclotron resonances, corresponding to gyroresonance numbers $s=0$ and $\pm1$, respectively. In the absence of streaming the response tensor for various specific 1D relativistic distributions, including a J\"uttner distribution, is known \citep{MGKF99,RMM1}, with the Cerenkov and cyclotron resonances contributing terms that involve relativistic plasma dispersion functions (RPDFs), $W(z)$ and $R(z - sy/\gamma)$, $S(z - sy/\gamma)$, respectively, where $ \beta = z $ and $ \beta = z - sy/\gamma$ satisfy the Cerenkov and cyclotron resonances, respectively. In the low-frequency approximation $x=\omega/\Omega_e\ll1$, the contribution of the cyclotron resonances to the response tensor may be neglected. In this case, the allowed wave modes of a pulsar plasma are referred to as the X-mode and the O-mode, which were originally defined assuming a cold-plasma model for the distribution function \citep{AB86,BA86}. These modes are linearly polarized; the  O-mode has a longitudinal component and is referred to as the LO-mode when this is taken into account. The cyclotron-resonant contributions  lead to elliptically polarized wave modes, and are important when discussing effects associated with observed circularly- or elliptically-polarized components of pulsar radio emission.

One motivation for our derivation of the general form of the dielectric tensor here is to discuss the polarization changes as radio waves escape from pulsars or magnetars and encounter the cyclotron resonance region. We refer to the polarization changes as generalized Faraday rotation (GFR), also called Faraday conversion. Astrophysical applications of GFR have been suggested in connection with the circularly-polarized component in synchrotron sources \citep[e.g.,][]{2011MNRAS.416.2574H}, the circularly-polarized component in pulsar radio emission  \citep{1979AuJPh..32...61M,1998Ap&SS.262..379L,2006MNRAS.366.1539P,2010MNRAS.403..569W,2012MNRAS.425..814B}, and with more complicated features of the polarization of radio emission from magnetars \citep[e.g.,][]{2007MNRAS.377..107K} and fast radio  bursts (FRBs) \citep[e.g.,][]{2019MNRAS.485L..78V,2019ApJ...876...74G}. In most of these discussions, the plasma is assumed to be a (cold) magnetoionic medium.

The general form of the response tensor for a pulsar plasma is cumbersome, and we discuss relevant approximations to it. We comment on the cold-plasma (magnetoionic) limit often assumed in discussions of GFR, and note that it is a poor approximation for the plasma around pulsars, magnetars and FRBs. The weak-anisotropy approximation (WAA) is a useful approximation to the general form of the response tensor and particularly relevant to GFR. In the WAA the waves are assumed to be transverse waves in vacuo to zeroth order in an expansion in the components of the dielectric tensor. In the WAA, both the refractive indices and polarization vectors of the two natural wave modes are determined to first order in this expansion. The polarization changes implied by GFR may be described in terms of the motion  on the Poincar\'e sphere of a  point $P$ representing the polarization of the wave. Conjugate points on the Poincar\'e sphere that represent the orthogonal polarizations of the two wave modes (in the WAA) define an axis, referred to here as the GFR axis, about which $P$ rotates as the wave propagates. As a wave with $x=\omega/\Omega_e\ll1$ at the emission point propagates away from the star, $\Omega_e$ decreases, implying that $x$ increases. The cyclotron resonance would be encountered at $x=1$ in a cold non-streaming plasma.  The inclusion of relativistic streaming, $\gamma_{\rm s}\gg1$, lowers the frequency at which the cyclotron resonance is encountered by a factor of order of the Lorentz factor of the streaming, and inclusion of a relativistic spread, $\aV{\gamma}\gg1$, with $ \aV{Q} $ denoting the average of $ Q $, in energies in plasma rest frame smooths the cyclotron resonance over a range of frequencies. It is important to include the intrinsic spread in $\gamma$ in any quantitative theory for GFR \citep{2004PhRvE..70a6404M,2004IAUS..218..381L}.

\section{Dielectric tensor for a pulsar plasma}
\label{sect:Kij}

The dielectric tensor for a pulsar plasma  may be deduced from the general forms for the response tensor $K_{ij} (\omega,\bm{k})$, which are derived using plasma kinetic theory. In this section, two general forms for $K_{ij} (\omega,\bm{k})$, derived using the Vlasov method and the forward-scattering method \citep{M13}, are written down and the 1D approximation is made to them.

\subsection{Vlasov form of the response tensor}

The Vlasov method gives the response tensor $K_{ij} (\omega,\bm{k}) $ as
\begin{multline}\label{eq:Kij_Vlasov}
	K_{ij}(\omega,\bm{k})
	= \delta_{ij} + \sum \frac{q^2}{\varepsilon_0 \omega^2} \int d^3\bm{p}
	\bigg[
		b_i b_j \frac{v_\parallel}{v_\perp} \left(v_\perp \frac{\partial}{\partial p_\parallel} - v_\parallel \frac{\partial}{\partial p_\perp}\right)
		\\
		+ \sum_{s=-\infty}^\infty \frac{V_i(\bm{k},\bm{p};s)V_j^*(\bm{k},\bm{p};s)}{\omega-s\Omega-k_\parallel v_\parallel} \left(\frac{\omega- k_\parallel v_\parallel}{v_\perp} \frac{\partial}{\partial p_\perp} + k_\parallel \frac{\partial}{\partial p_\parallel}\right)
		\bigg] f(\bm{p}),
\end{multline}
where $ \delta_{ij} $ is the Kronecker delta; the unlabeled sum is over all (unlabelled) species of particles with charge $q$, mass $m$, relativistic gyrofrequency $ \Omega = \Omega_e / \gamma $, distribution function $f(\bm{p})$, velocity $ \bm{v} $ and 3-momentum $ \bm{p} = \gamma m \bm{v} $; the parallel and perpendicular components, with respect to unit vector $ \bm{b} $ along the the magnetic field, of vector quantities are denoted respectively by subscripts $ \parallel $ and $ \perp $; and
\begin{equation}\label{eq:Vkps_Vlasov}
	\bm{V}(\bm{k},\bm{p};s) =
	\left(
	v_\perp \frac{s}{k_\perp R} J_s(k_\perp R),
	-i \epsilon v_\perp J'_s(k_\perp R),
	v_\parallel J_s(k_\perp R)
	\right),
\end{equation}
where $ J_s (k_\perp R) $ is a Bessel function of the first kind with argument $k_\perp R$, where $R=p_\perp/|q|B$ is the radius of gyration, and $ J'(x) = d J(x) / d x $. Here it is assumed that there are only two species, electrons and positrons labeled $\epsilon=\mp$, respectively.

\subsubsection{Antihermitian part of $K_{ij}(\omega,\bm{k})$}

The antihermitian part of \eqref{eq:Kij_Vlasov} is given by
\begin{multline}\label{eq:Kij_Vlasov_Antihermitian}
	K^A_{ij}(\omega,\bm{k})
	= - \sum{i\pi q^2\over\varepsilon_0\omega^2}\int d^3\bm{p} \sum_{s=-\infty}^\infty V_i(\bm{k},\bm{p};s)V_j^*(\bm{k},\bm{p};s) \delta(\omega - s\Omega_e / \gamma -k_\parallel v_\parallel)\\
	\times
	\left(
	\frac{s\Omega_e}{ \gamma v_\perp} \frac{\partial}{\partial p_\perp} +k_\parallel \frac{\partial}{\partial p_\parallel}
	\right)f(\bm{p})
\end{multline}
and describes dissipation due to gyroresonant interactions satisfying
\begin{equation}
	\omega-s\Omega_e/\gamma-k_\parallel v_\parallel=0,
	\label{gyrores1}
\end{equation}
where the gryroresonant number, $s$, is an integer. For a 1D distribution only $s=0,\pm1$ contribute.

\subsubsection{1D assumption}

In the 1D case, the perpendicular momentum, $p_\perp$, of all particles is identically zero, and the 3D distribution function $f(\bm{p})$ may be replaced by the 1D distribution function $g(u)$, where $u=p_\parallel/mc=\gamma\beta$ is the 4-speed (in units of $c$). Due to  $v_\perp=0$, the argument of the Bessel functions in \eqref{eq:Vkps_Vlasov} is zero, and one has
\begin{equation}
	\bm{V}(\bm{k},\bm{p};0)=v_\parallel \bm{b},
	\quad
	\bm{V}(\bm{k},\bm{p};\pm1)=\half v_\perp\bzeta(\pm1),
	\quad
	\bzeta(s)=(1,-is\epsilon,0),
	\label {bzetas}
\end{equation}
with $\bm{V}$ being zero for all $|s|>1$ and where we choose $ \bm{b} = (0, 0, 1) $ so that the magnetic field is along the 3-axis. Note that although the terms $\bm{V}(\bm{k},\bm{p};\pm1)$ are zero for $v_\perp=0$ in the 1D case, these terms need to be retained in the Vlasov form \eqref{eq:Kij_Vlasov}, because the $p_\perp$-derivative  in \eqref{eq:Kij_Vlasov_Antihermitian} acts on $\delta(p_\perp)$ and hence one needs to partially integrate this term to evaluate it. One sets $v_\perp=0$ only after this partial integration. One has, for $s=\pm1$,
\begin{multline}
	\int\d^3\bm{p}V_i(\bm{k},\bm{p};s)V_j^*(\bm{k},\bm{p};s) \delta\left(\omega - s\Omega_e/\gamma - k_\parallel v_\parallel\right) \frac{s\Omega_e}{\gamma v_\perp}\frac{\partial}{\partial p_\perp} \left[\frac{\delta(p_\perp)}{2\pi p_\perp}g(u)\right]\\
	=
	-\frac{sy}{2m}\,\zeta_i(s)\zeta_j^*(s)\int\d u\,\delta\left(u + s y - \gamma z\right)g(u),
	\label{ppd1}
\end{multline}
with  $z=\omega/k_\parallel c$ and $y=\Omega_e/k_\parallel c$.

\subsection{Forward-scattering form for the dielectric tensor}

An alternative general form for the dielectric tensor $ K_{ij}(\omega,\bm{k}) $ is obtained using the forward-scattering method \citep{Melrose87c}. This form is given by
\begin{multline} \label{eq:KijmagF}
	K_{ij}(\omega,\bm{k})
	= \delta_{ij} - \sum\sum_{s=-\infty}^\infty{q^2\over\varepsilon_0m\omega^2}
	\int \d^3\bm{p}\, \frac{f(\bm{p})}{\gamma}\bigg\{
	J_s^2(k_\perp R) \tau_{ij}(\omega_s)\\
	+ \frac{J_s(k_\perp R)}{\omega_s}\left[
		\tau_{im}(\omega_s) k_m V_j^*(\bm{k},\bm{p};s)
		+V_i(\bm{k},\bm{p};s) k_l \tau_{lj}(\omega_s)
		\right]\\
	+ \frac{1}{\omega_s^2}\left[k_l k_m \tau_{lm}(\omega_s) - \frac{\omega^2}{c^2}
		\right] V_i(\bm{k},\bm{p};s) V_j^*(\bm{k},\bm{p};s)
	\bigg\},
\end{multline}
where repeated subscripts $l$ and $m$ imply sums from 1 to 3 and with $ \omega_s = \omega-s\Omega_e/\gamma-k_\parallel v_\parallel $,
\begin{equation}
	\tau_{ij}(\omega_s)
	=
	\begin{pmatrix}
		\dfrac{\omega_s^2}{\omega_s^2-\Omega^2}               &
		\dfrac{i\epsilon\omega_s\Omega}{\omega_s^2-\Omega^2}  & 0     \\
		\dfrac{-i\epsilon\omega_s\Omega}{\omega_s^2-\Omega^2} &
		\dfrac{\omega_s^2}{\omega_s^2-\Omega^2}               & 0     \\
		0                                                 & 0 & 1
	\end{pmatrix}.
	\label{tauij8}
\end{equation}

As in \eqref{eq:Kij_Vlasov}, the sum in \eqref{eq:KijmagF} is over all unlabeled species, with only electrons and positrons, $\epsilon=\mp$, relevant here.
The form \eqref{eq:KijmagF} may also be obtained from \eqref{eq:Kij_Vlasov} by a tedious calculation involving partially integrating and using recursion relations and sum rules for the Bessel functions.

\subsection{Averages over a 1D distribution}

Consider a pulsar plasma, composed of electrons, $\epsilon=-$, and positrons, $\epsilon=+$. The average of an arbitrary function $Q$ over the 1D distribution function, $\tr{g}{\epsilon}(u)$, for electrons or positrons is given by
\begin{equation}
	\tR{n}{\epsilon} \tr{\aV{Q}}{\epsilon}
	= \int du\,Q\,\tr{g}{\epsilon}(u)
	= \int d\beta\, \gamma^3 Q\, \tr{g}{\epsilon}(u),
	\label{eq:average}
\end{equation}
which defines the number density, $\tR{n}{\epsilon}$, for $Q=1$.\footnote{The number density is written $\tR{n}{\epsilon}$ to avoid confusion with the refractive indices, $n_\pm$, of the two orthogonal modes.} The average of $ Q $ over the combined distribution $ g(u) = \tr{g}{+}(u) + \tr{g}{-}(u) $ is given by $ n \aV{Q} = \tR{n}{+} \tR{\aV{Q}}{+} + \tR{n}{-} \tR{\aV{Q}}{-} $ with combined number density $ n = \tR{n}{+} + \tR{n}{-} $. When electron and positron number densities are equal we have $ \tR{n}{+} = \tR{n}{-} = n / 2 $. The charge and current densities are then given by
\begin{equation}
    \eta
        = \sum_{\epsilon} \epsilon e \tR{n}{\epsilon},\quad
    \bm{J}
        = c \sum_{\epsilon} \epsilon e \tR{n}{\epsilon} \tR{\aV{\beta}}{\epsilon}.
\end{equation}

Averages over primed quantities in the primed frame can be somewhat counter-intuitive. For example, setting $X=1$, the average 
$\aV{1/\gamma'}^{{\epsilon}{\prime}} = \gamma_{\rm s}^{-1} \aV{1/\gamma}^{\epsilon} $ 
is much narrower than the average ${\aV{1/\gamma}}^{\epsilon} $ in the unprimed frame, and, setting $X=\gamma'^2$, the average of $\gamma'$ in the primed frame is much larger than the average of $\gamma$ in the unprimed frame, specifically, $\aV{\gamma'}^{{\epsilon}{\prime}} = \gamma_{\rm s} \tR{\aV{\gamma(1+\beta^2\beta_{\rm s}^2}}{\epsilon} \approx 2\gamma_{\rm s} \tR{\aV{\gamma}}{\epsilon} $ for $\gamma,\gamma_{\rm s}\gg1$.

\subsection{Response tensor for a 1D distribution}

In the 1D case, the only contributions to the sum over $s$ in \eqref{eq:KijmagF} are for $s=0,\pm1$, corresponding to resonant denominators $1/\omega_0^2$ and $1/(\omega_0^2-\Omega^2)$, with $\omega_0=\omega-k_\parallel v_\parallel=\omega(z-\beta)/z$ and $\Omega=\Omega_e/\gamma=\omega y/z\gamma$.  Assuming, without loss of generality, the magnetic field along the 3-axis and the wave vector in the 1-3 plane, one has
\begin{equation}
	\bm{k}=(k_\perp,0,k_\parallel )=\frac{\omega}{zc}(\tan\theta,0,1).
	\label{bik}
\end{equation}
With this notation, the dielectric tensor \eqref{eq:KijmagF} for a pulsar plasma may be written in the form
\begin{equation}\label{eq:Kij1_and_NPP1}
    K_{ij}(\omega,\bm{k})
        =\delta_{ij}+\frac{\Pi_{ij}(\omega,\bm{k})}{\varepsilon_0\omega^2},
    \quad\text{with}\quad
    \Pi_{ij}(\omega,\bm{k})
        =-\sum_{\epsilon}\frac{e^2\tR{n}{\epsilon}}{m} \tR{\aV{\frac{A_{ij}(\omega,\bm{k};\beta)}{\gamma}}}{\epsilon},
\end{equation}
with $A_{ij}(\omega,\bm{k};\beta)\to A_{ij}$ given by
\begin{equation}\label{eq:Aij1}
\begin{gathered}
    A_{11}
        = A_{22}
        = \frac{\omega_0^2}{\omega_0^2-\Omega^2},\quad
    A_{33}
        = \frac{\omega^2}{\gamma^2\omega_0^2} + \frac{\omega^2}{\omega_0^2-\Omega^2}\left(\frac{\beta\tan\theta}{z}\right)^2,\\
    A_{12}
        = -A_{21}
        = \rmi\epsilon \frac{\omega_0\Omega}{\omega_0^2-\Omega^2},\quad
    A_{13}
        = A_{31} 
        = \frac{\omega_0\omega}{\omega_0^2-\Omega^2}\,\left(\frac{\beta\tan\theta}{z}\right),\\
    A_{23}
        = -A_{32}
        = -\rmi\epsilon	\frac{\omega\Omega}{\omega_0^2-\Omega^2}\,\left(\frac{\beta\tan\theta}{z}\right).
\end{gathered}
\end{equation}
Writing the dielectric tensor in the form \eqref{eq:Kij1_and_NPP1} facilitates Lorentz-transforming it, due to the tensor $\Pi_{ij}$ being the space components of a 4-tensor, as discussed in \S\ref{sect:LTKij}.

\subsection{Response tensor for a cold pair plasma}
\label{sect:cold_stream}

If the spread in energy in the pair plasma is neglected, the resulting model corresponds to a cold streaming pair plasma. In the rest frame $ \mathcal{K} $, the response tensor for such a model follows by making the replacement $ du \tr{g}{\epsilon}(u) \to du \tR{n}{\epsilon} \delta(u) $ when evaluating the average in equation \eqref{eq:Kij1_and_NPP1}. Assuming that electron and positron distributions have a common rest frame, $ \mathcal{K} $, we obtain
\begin{equation}
	\Pi_{ij}(\omega,\bm{k})
        =-\sum_{\epsilon}\frac{e^2\tR{n}{\epsilon}}{m} \left. \tau_{ij}\left(\omega_s\right)\right|_{\beta = 0 }.
	\label{NPP1_cold}
\end{equation}

\section{Solutions of resonance conditions}
\label{sect:resonances}

The dielectric tensor for a magnetized plasma has resonances when the gyroresonance condition
\begin{equation}
	\omega - s\Omega - k_\parallel v_\parallel = 0,
	\quad \text{or} \quad
	z - sy/\gamma - \beta =0,
	\label{gyroresonance}
\end{equation}
 is satisfied, where $s$ is the harmonic number and $\Omega=\Omega_e/\gamma$ is the relativistic gyrofrequency,  and in the second form, $z=\omega/k_\parallel c$, $y=\Omega_e/k_\parallel c$. 
In a 1D electron gas,  the perpendicular component of the velocity is zero, $v_\perp=0$, and only the Cerenkov resonance $s=0$ and the cyclotron resonances $s=\pm1$ are allowed. The Cerenkov resonance is at $\beta=z$. The two cyclotron resonances are at solutions of
\begin{equation}
	(\beta-z)^2=y^2(1-\beta^2),
	\label{gyroresa}
\end{equation}
which is a quadratic equation for $\beta$. The two solutions, $\beta=\beta_\pm$ say, of the quadratic equation \eqref{gyroresa} for the cyclotron resonances are
\begin{equation}\label{eq:beta_pm}
	\beta_\pm=\frac{z\pm \abs{y}(1+y^2-z^2)^{1/2}}{1+y^2}.
\end{equation}
From the resonance condition \eqref{gyroresonance} we have $z-\beta_\pm=sy/\gamma_\pm$ which implies that
\begin{equation}
	\gamma_\pm
	= s \sgn(y) \frac{1 + y^2}{\abs{y} z \mp \left(1 + y^2 - z^2\right)^{1/2}},
	\qquad
	u_\pm
	= s \sgn(y) \frac{z\pm \abs{y}(1+y^2-z^2)^{1/2}}{\abs{y} z \mp \left(1 + y^2 - z^2\right)^{1/2}},
	\label{upm}
\end{equation}
where $u_\pm=\gamma_\pm\beta_\pm$ is the corresponding 4-speed. We may also write \eqref{upm} as
\begin{equation}
	\gamma_\pm
	= s \sgn(y) \frac{\abs{y} z \pm \left(1 + y^2 - z^2\right)^{1/2}}{z^2 - 1},
	\qquad
	u_\pm
	= s \sgn(y) \frac{\abs{y} \pm z (1+y^2-z^2)^{1/2}}{z^2 - 1}.
	\label{upm2}
\end{equation}

\subsection{Interpretation of $ \pm $-solutions}

In interpreting the $ \pm $-solutions $\beta=\beta_\pm$ and $\gamma=\gamma_\pm$,  first consider the symmetry properties between upgoing and downgoing waves. Let upgoing (downgoing) waves and particles be identified as $k_\parallel>0$ ($k_\parallel<0$)  and $\beta>0$ ($\beta<0$), respectively. The transformation $z,y\to-z,-y$ interchanges the role of upgoing and downgoing waves and particles. The solutions $\beta_\pm$ reverse sign under $z,y\to-z,-y$, whereas $\gamma_\pm$ and the resonance conditions $z-\beta_\pm=sy/\gamma_\pm$ are unchanged. In the following discussion $z,y>0$ is assumed with solutions for $z,y<0$ following from these symmetry properties.

\begin{figure}
	\centering
	\includegraphics[width=\textwidth]{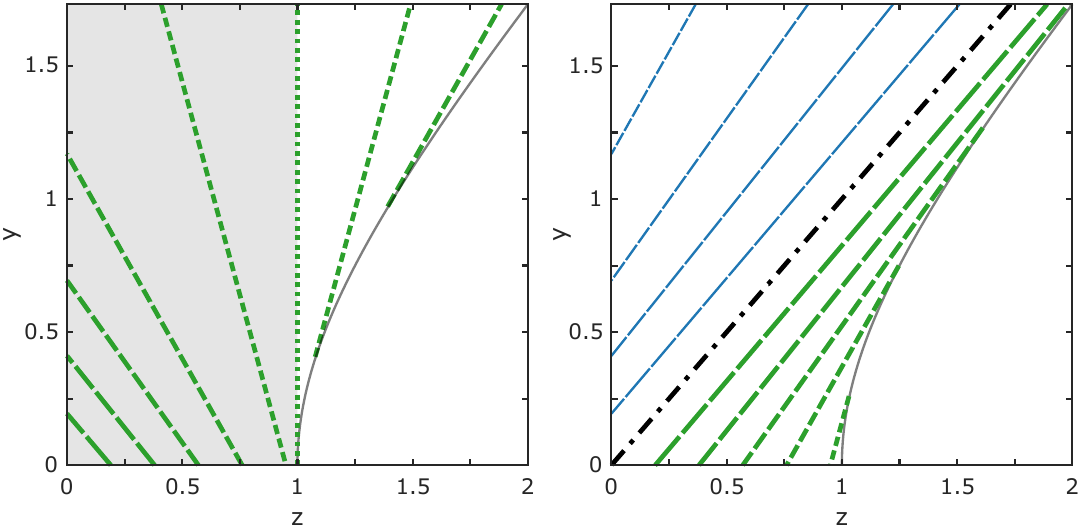}
	\caption{Contour plots of $ \beta_{+} $ (left panel) and $ \beta_{-} $ (right panel) as a function of $ y $ and $ z $ for $ \lvert\beta_{\pm}\rvert = 1 $ (vertical dotted), 0 (dash-dotted), and $ 0.19, 0.38, 0.57, 0.76, 0.95 $ (decreasing dash length). The negative contours are in thin blue and positive contours are in thick green. The thin grey curves indicate the line $ 1 + y^2 - z^2 = 0 $. The shaded regions correspond to anomalous Doppler effect.}
	\label{fig:beta_pm_contours_yz}
\end{figure}

Figure~\ref{fig:beta_pm_contours_yz} shows contour plots of $ \beta_{+} $ (left panel) and $ \beta_{-} $ (right panel) as a function of $ y $ and $ z $ for $ \lvert\beta_{\pm}\rvert = 1 $ (vertical dotted), 0 (dash-dotted), and $ 0.19, 0.38, 0.57, 0.76, 0.95 $ (decreasing dash length). The negative contours are in thin blue and positive contours are in thick green. The thin grey curves indicate the line $ 1 + y^2 - z^2 = 0 $.

The $ \pm $-solutions in equations \eqref{eq:beta_pm} are real only for $z^2 \leq 1+y^2$. Equations~\eqref{upm} imply that the anomalous Doppler resonance, $s=-1$, may be satisfied only for the $ \beta = \beta_+ $ solution over $ 0 \leq z < 1 $, which is shown as a shaded region in Figure~\ref{fig:beta_pm_contours_yz}; and the Doppler resonance, $ s = +1 $, may be satisfied for the $ \beta = \beta_+ $ solution over $ 1 < z < \sqrt{1+y^2} $, and for the $ \beta = \beta_- $ solution over $ 0 \leq z < \sqrt{1+y^2} $. For the $ + $-solution the transition between solutions for $s=+1$ and $s=-1$ occurs for $\gamma_+=\infty$, $\beta_+=+1$ (vertical dotted green line). It follows that for the normal Doppler effect, $s=+1$, there is a solution $\beta=\beta_-$ for subluminal waves, $z<1$, and two solutions,  $\beta=\beta_+, \beta=\beta_-$ for superluminal waves, $ 1 \leq z < \sqrt{1+y^2} $. The anomalous Doppler effect, $s=-1$, requires subluminal waves, $z<1$, and then only for $\beta=\beta_+$.

\subsection{Plots of  $\beta_\pm$ as functions of $z$ for fixed $x=\omega/\Omega_e$}
The resonant solutions,  $\beta_\pm$, as functions of $z$ for fixed $x=\omega/\Omega_e$ are of specific interest in the low-frequency limit, $x\ll1$, and around the cyclotron resonance, where $x$ is of order unity. The dependence on $x$ may be shown by writing \eqref{eq:beta_pm} and \eqref{upm} in the form
\begin{equation}\label{betazx}
    \beta_\pm
        = \frac{zx^2\pm \abs{z} \left[x^2(1-z^2)+z^2\right]^{1/2}}{x^2+z^2},\quad
    \gamma_\pm
         = s \sgn{(z)} \frac{z \abs{z} \pm \left[x^2(1-z^2)+z^2\right]^{1/2}}{x(z^2 - 1)}.
\end{equation}

\begin{figure}
	\centering
	\includegraphics[width=\textwidth]{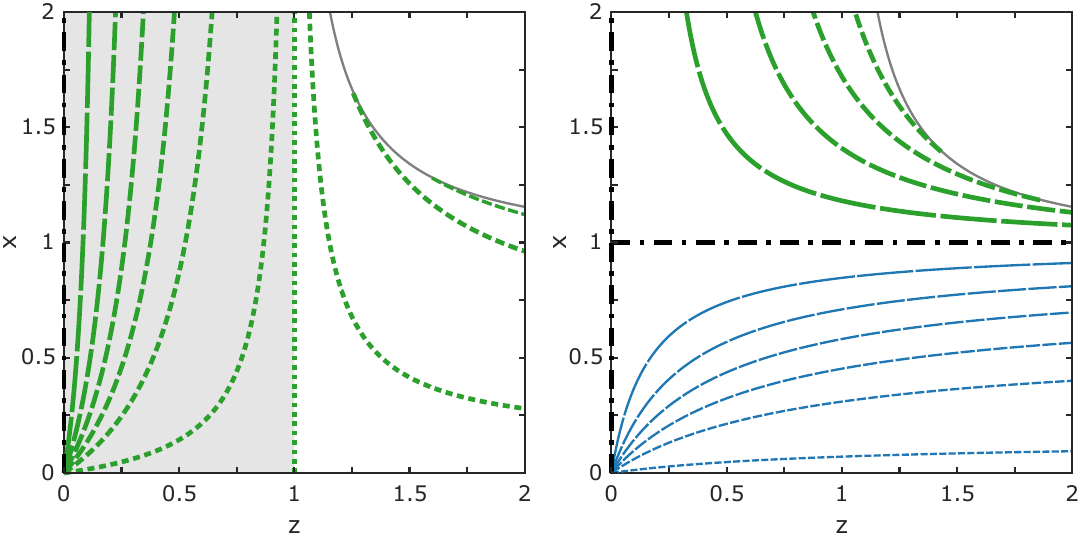}
	\caption{Contour plots of $ \beta_{+} $ (left panel) and $ \beta_{-} $ (right panel) as a function of $ x $ and $ z $ for $ \lvert\beta_{\pm}\rvert = 1 $ (vertical dotted), 0 (dash-dotted), and $ 0.17, 0.33, 0.5, 0.66, 0.83, 0.99 $ (decreasing dash length). The negative contours are in thin blue and positive contours are in thick green. The thin grey curves indicate the line $ x^2 (1 - z^2) + z^2 = 0 $. The shaded region corresponds to anomalous Doppler effect.}
	\label{fig:beta_pm_contours_xz}
\end{figure}

Figure~\ref{fig:beta_pm_contours_xz} shows contour plots of $ \beta_{+} $ (left panel) and $ \beta_{-} $ (right panel) as a function of $ x $ and $ z $ for $ \lvert\beta_{\pm}\rvert = 1 $ (vertical dotted), 0 (dash-dotted), and $ 0.17, 0.33, 0.5, 0.66, 0.83, 0.99 $ (decreasing dash length). The negative contours are in thin blue and positive contours are in thick green. The thin grey curves indicate the line $ x^2 (1 - z^2) + z^2 = 0 $. The shaded region corresponds to the anomalous Doppler effect.

\begin{figure}
	\centering
	\includegraphics[width=\textwidth]{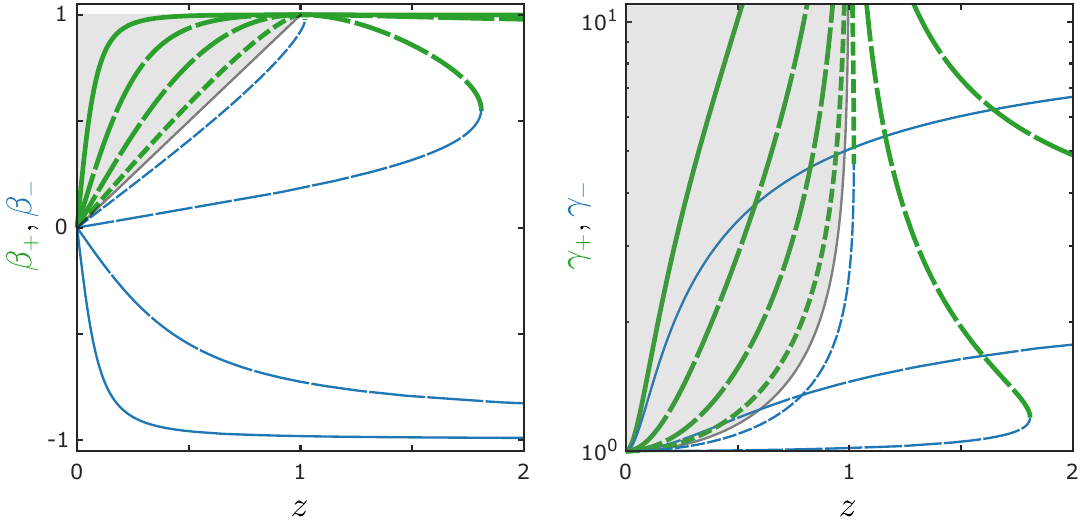}
	\caption{Contour plots of $ \beta_{\pm} $ (left panel) and corresponding $ \gamma_{\pm} $ (right panel) as a function of $ z $ for $ x = 0.1 $ (solid) and $ 0.4, 1.2, 5 $ (decreasing dash length). Contours of $ \beta_+, \gamma_+ $ are in thick green and those of $ \beta_-, \gamma_- $ are in thin blue. The curves $\beta_+$ and $\beta_-$ extend to $z=\infty$ for $x<1$, and they form single closed curves for $x>1$ meeting at $ z^2 = x^2 / (x^2 - 1) $. For $x\gg1$, illustrated by $x=5$, the closed curve approaches the line $\beta = z$ (thin grey)  for $z\le1$, which corresponds to the Cerenkov resonance. The shaded regions correspond to the anomalous Doppler effect.}
	\label{fig:betapm}
\end{figure}

Figure~\ref{fig:betapm} shows plots of $ \beta_{\pm} $ (left panel) and corresponding $ \gamma_{\pm} $ (right panel) as a function of $ z $ for $ x = 0.1 $ (solid) and $ 0.4, 1.2, 5 $ (decreasing dash length). Contours of $ \beta_+, \gamma_+ $ are in thick green and those of $ \beta_-, \gamma_- $ are in thin blue. The curves $\beta_+$ and $\beta_-$ extend to $z=\infty$ for $x<1$, and they form single closed curves for $x>1$ meeting at $ \lvert z\rvert = x / (x^2 - 1)^{1/2} $. For $x\gg1$, illustrated by $x=5$, the closed curve approaches the line $\beta = z$  for $z\le1$, which corresponds to the Cerenkov resonance. The shaded regions correspond to anomalous Doppler effect; which are bounded on the right by $ \beta_\pm = z $ and $ \gamma_\pm = (1 - z^2)^{-1/2} $ which are the limits of $ \beta_\pm $ and $ \gamma_\pm $ as $ x \to \infty $. In the low-frequency limit, $x\ll1$, the solutions may be approximated by
\begin{equation}
	\beta_\pm
	=
	\begin{dcases}
		\frac{\pm \abs{z}}{\left(x^2 + z^2\right)^{1/2}}, & \text{if}\quad \abs{z} \ll 1,     \\
		\pm \left(2 + z\right) \abs{z},                   & \text{if}\quad \abs{z + 1} \ll 1, \\
		\pm \left(2 - z\right) \abs{z},                   & \text{if}\quad \abs{z - 1} \ll 1,
	\end{dcases}
\end{equation}
with corresponding Lorentz factors
\begin{equation}
	\gamma_\pm
	=
	\begin{dcases}
		- s \frac{z \abs{z} \pm \left(x^2 + z^2\right)^{1/2}}{x},                      & \text{if}\quad \abs{z} \ll 1,     \\
		s \frac{\left(z - 1\right) \left(\abs{z} \mp 1\right)}{4x \left(z + 1\right)}, & \text{if}\quad \abs{z + 1} \ll 1, \\
		s \frac{\left(z + 1\right) \left(\abs{z} \pm 1\right)}{4x \left(z - 1\right)}, & \text{if}\quad \abs{z - 1} \ll 1,
	\end{dcases}
\end{equation}
and there is no restriction on $z$. For $x>1$ real solutions exist only for $z^2\le x^2/(x^2-1)$, with the curves for the $\pm$-solutions joining smoothly at $z^2=x^2/(x^2-1)$.

\subsection{Three RPDFs}
\label{ThreeRPDFs}

We may write $ A_{ij}$, using \eqref{eq:Aij1} and \eqref{eq:beta_pm}, as
\begin{align}
	A_{11}
        & = A_{22} 
        = \frac{(z-\beta)^2}{(1+y^2)(\beta-\beta_+)(\beta-\beta_-)},\nn\\
	A_{33}
        & = \frac{z^2}{\gamma^2(z-\beta)^2} + \frac{\beta^2\tan^2\theta}{(1+y^2)(\beta-\beta_+)(\beta-\beta_-)},\nn\\
	A_{13}
        & =A_{31}
        = \frac{(z-\beta)\beta\tan\theta}{(1+y^2)(\beta-\beta_+)(\beta-\beta_-)},\nn\\
	A_{12}
        & = -A_{21}
        = \rmi\epsilon \frac{y(z-\beta)}{\gamma(1+y^2)(\beta-\beta_+)(\beta-\beta_-)},\nn\\
	A_{23}
        & = -A_{32}
        = - \rmi\epsilon \frac{y\beta\tan\theta}{\gamma(1+y^2)(\beta-\beta_+)(\beta-\beta_-)}.
	\label{eq:Aij2}
\end{align}
Writing
\begin{equation}
    \frac{1}{(\beta-\beta_+)(\beta-\beta_-)}
    \begin{bmatrix}
        1 \\
        \beta\\
        \beta^2\\
        z-\beta\\
        (z-\beta)^2\\
        (z-\beta)\beta
    \end{bmatrix}
    = 
    \begin{bmatrix}
        0\\
        0\\
        1\\
        0\\
        1\\
        -1
    \end{bmatrix}
	+ \frac{1}{\beta_+ - \beta_-} \sum_{\alpha=\pm}\frac{\alpha}{\beta-\beta_\alpha}
    \begin{bmatrix}
        1 \\
        \beta_\alpha\\
        \beta_\alpha^2\\
        z-\beta_\alpha\\
        (z-\beta_\alpha)^2\\
        (z-\beta_\alpha)\beta_\alpha
	\end{bmatrix},
	\label{ur1}
\end{equation}
allows us to write the averages in \eqref{eq:Kij1_and_NPP1} in terms of the three RPDFs:
\begin{equation}
	\tr{W}{\epsilon}(z)=\tr{\aV{\frac{1}{\gamma^3(\beta-z)^2}}}{\epsilon},
	\quad
	\tr{R}{\epsilon}(\beta_\alpha)=
	\tr{\aV{\frac{1}{\gamma(\beta-\beta_\alpha)}}}{\epsilon},
	\quad
	\tr{S}{\epsilon}(\beta_\alpha)=\tr{\aV{\frac{1}{\gamma^2(\beta-\beta_\alpha)}}}{\epsilon},
	\label{WRS1}
\end{equation}
with the averages to be understood as defined by \eqref{eq:average}. The RPDF $ \tr{W}{\epsilon}(z) $ arises from the Cerenkov resonance, and RPDFs $ \tr{R}{\epsilon}(\beta_\alpha) $ and $ \tr{S}{\epsilon}(\beta_\alpha) $ arise from the cyclotron resonances, $\beta=\beta_\alpha$. An alternative form for $\tr{W}{\epsilon}(z)$ is
\begin{equation}
	\tr{W}{\epsilon}(z)=\frac{1}{\tR{n}{\epsilon}}\int\rmd \beta\frac{1}{\beta-z-\rmi0}\frac{\rmd \tr{g}{\epsilon}(u)}{\rmd \beta},
	\label{Wze}
\end{equation}
which may be written as, \citep{RMM1},
\begin{equation}
	\tr{W}{\epsilon}(z)
	=
	\begin{dcases}
		\tr{\aV{\frac{1}{\gamma^3(\beta-z)^2}}}{\epsilon},                                                                                                                                                                   & \quad \text{for } \abs{z} > 1,    \\
		\frac{1}{\tR{n}{\epsilon}}\left[i \pi \left.\frac{\rmd \tr{g}{\epsilon}(u)}{\rmd \beta}\right\vert_{\beta = z} + \wp \int \rmd \beta\, \frac{1}{\beta - z} \frac{\rmd \tr{g}{\epsilon}(u)}{\rmd \beta}\right], & \quad \text{for } \abs{z} \leq 1,
	\end{dcases}
\end{equation}
where $ \wp $ indicates a Cauchy principle value integral. The imaginary parts of these RPDFs follow using the Landau prescription, which gives the term $-\rmi0$ in \eqref{Wze}; the singularity contributes a semi-residue, found by replacing the resonant denominator by $+i\pi\delta(\beta-z)$ in the numerator.  The resonant parts of $\tr{R}{\epsilon}(\beta_\alpha)$ and $\tr{S}{\epsilon}(\beta_\alpha)$ follow by replacing $1/(\beta-\beta_\alpha)$ by $1/(\beta-\beta_\alpha-\rmi0)$,
\begin{equation}
	\tr{R}{\epsilon}(\beta_\alpha)=\frac{1}{\tR{n}{\epsilon}}\int\rmd \beta\,\frac{\gamma^2}{\beta-\beta_\alpha-\rmi0} \tr{g}{\epsilon}(u),
	\qquad
	\tr{S}{\epsilon}(\beta_\alpha)=\frac{1}{\tR{n}{\epsilon}}\int\rmd \beta\,\frac{\gamma}{\beta-\beta_\alpha-\rmi0} \tr{g}{\epsilon}(u),
	\label{ImRS}
\end{equation}
with the semi-residues giving the imaginary parts. We note that $ \abs{\beta_\alpha} < 1 $ which implies that
\begin{align}
	\tr{R}{\epsilon}(\beta_\alpha)
	 & = \frac{1}{\tR{n}{\epsilon}}\left[i \pi \left.\gamma^2 \tr{g}{\epsilon}(u)\right\vert_{\beta = \beta_\alpha} + \wp \int \rmd \beta\, \frac{\gamma^2}{\beta - \beta_\alpha} \tr{g}{\epsilon}(u)\right], \\
	\tr{S}{\epsilon}(\beta_\alpha)
	 & = \frac{1}{\tR{n}{\epsilon}}\left[i \pi \left.\gamma \tr{g}{\epsilon}(u)\right\vert_{\beta = \beta_\alpha} + \wp \int \rmd \beta\, \frac{\gamma}{\beta - \beta_\alpha} \tr{g}{\epsilon}(u)\right].
\end{align}

\begin{figure}
	\centering
	\includegraphics[width=\textwidth]{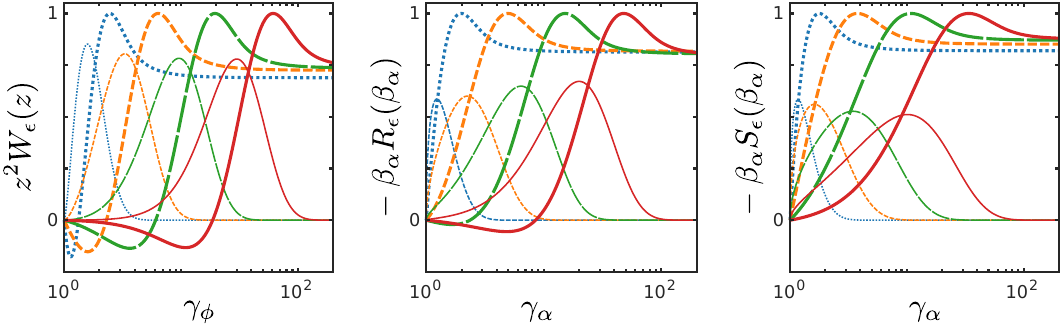}
	\caption{Plots of real (thick) and negative of the imaginary (thin) of RPDFs $ z^2 \tr{W}{\epsilon} (z) $ (left), $ - \beta_\alpha \tr{R}{\epsilon} (\beta_\alpha) $ (middle) and  $ - \beta_\alpha \tr{S}{\epsilon} (\beta_\alpha) $ (right) for $ \rho^\epsilon = 3.16 $ (blue, dotted), 1 (orange, short dashed), 0.316 (green, long dashed) and 0.1 (red, solid). The real parts of RPDFs are scaled to unity and the imaginary parts are scaled relative to the real parts.}
	\label{fig:WRS_z_beta_alpha}
\end{figure}

Figure~\ref{fig:WRS_z_beta_alpha} shows plots of real (top row) and imaginary (bottom row) of RPDFs $ \tr{W}{\epsilon} (z) $ (left column), $ \tr{R}{\epsilon} (\beta_\alpha) $ (middle column) and  $ \tr{S}{\epsilon} (\beta_\alpha) $ (right column) for $ \rho = 1 $ (solid), 5 (long dashed) and 25 (short dashed). The magnitudes of real and imaginary components of the RPDFs have been scaled to unity while preserving their signs.

\begin{figure}
	\centering
	\includegraphics[width=\textwidth]{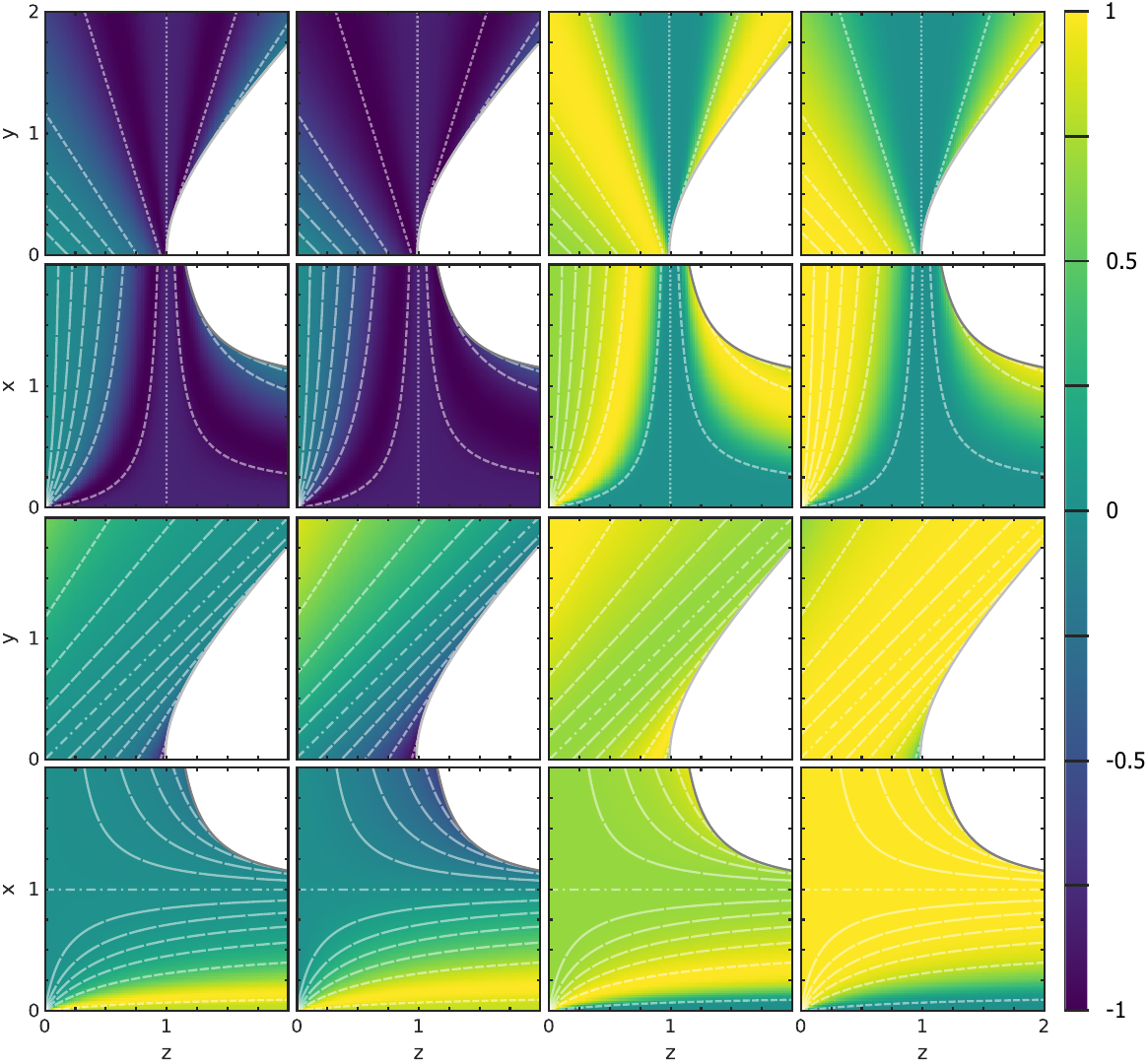}
	\caption{Colour plots of $ \Re \tr{R}{\epsilon} (\beta_\alpha) $ (first column), $ \Re \tr{S}{\epsilon} (\beta_\alpha) $ (second column), $ \Im \tr{R}{\epsilon} (\beta_\alpha) $ (third column), and $ \Im \tr{S}{\epsilon} (\beta_\alpha) $ (fourth column) for $ \alpha = + $ (first and second rows) and $ \alpha = - $ (third and fourth rows) with $ \beta_\alpha = \beta_\alpha(z, y) $ (first and third rows) and $ \beta_\alpha = \beta_\alpha(z, x) $ (second and fourth rows). We use $ \rho = 1 $ and contours of $ \beta_\alpha $ from Figures~\ref{fig:beta_pm_contours_yz} and \ref{fig:beta_pm_contours_xz} are superimposed (transparent white). The magnitudes of real and imaginary components of the RPDFs have been scaled to unity while preserving their signs.}
	\label{fig:RS_yz_xz}
\end{figure}

Figure~\ref{fig:RS_yz_xz} shows colour plots of $ \Re \tr{R}{\epsilon} (\beta_\alpha) $ (first column), $ \Re \tr{S}{\epsilon} (\beta_\alpha) $ (second column), $ \Im \tr{R}{\epsilon} (\beta_\alpha) $ (third column), and $ \Im \tr{S}{\epsilon} (\beta_\alpha) $ (fourth column) for $ \alpha = + $ (first and second rows) and $ \alpha = - $ (third and fourth rows) with $ \beta_\alpha = \beta_\alpha(z, y) $ (first and third rows) and $ \beta_\alpha = \beta_\alpha(z, x) $ (second and fourth rows). We use $ \rho = 1 $ and contours of $ \beta_\alpha $ from Figures~\ref{fig:beta_pm_contours_yz} and \ref{fig:beta_pm_contours_xz} are superimposed (transparent white). The magnitudes of real and imaginary components of the RPDFs have been scaled to unity while preserving their signs. We note that when $ \beta_\alpha = \beta_\alpha(z,y) $, a single point on the $ \left(\beta_\alpha, \tr{R}{\epsilon}(\beta_\alpha)\right) $ 2D curve maps to a 3D path in the $ \left(z, y, \tr{R}{\epsilon}(\beta_\alpha(z,y))\right) $ space. Projecting this path onto the $ \left(z, y\right) $ plane and indicating the corresponding value of $ \tr{R}{\epsilon}(\beta_\alpha(z,y)) $ by a colour-map results in mapping of a single point from $ \left(\beta_\alpha, \tr{R}{\epsilon}(\beta_\alpha)\right) $ to a curve in the $ \left(z, y\right) $ plane. These curves correspond to a single value of $ \beta_\alpha(z, y) $ and hence the value of $ \left(\beta_\alpha, \tr{R}{\epsilon}(\beta_\alpha)\right) $ is constant along these curves. The inclusion of contours of constant $ \beta_\alpha $ from Figures~\ref{fig:beta_pm_contours_yz} and \ref{fig:beta_pm_contours_xz} are meant to serve as aids. The 2D curve $ \left(\beta_\alpha, \tr{R}{\epsilon}(\beta_\alpha)\right) $ is thus mapped to 3D surface $ \left(z, y, \tr{R}{\epsilon}(\beta_\alpha(z,y))\right) $ (or to its projection onto $ \left(z, y\right) $ 2D plane). The same comments apply to $ \tr{S}{\epsilon}(\beta_\alpha) $ and also when $ \beta_\alpha = \beta_\alpha(z, x) $ is considered.

\subsection{$\Pi_{ij}(\omega,\vect{k})$ for a stationary J\"uttner distribution}
The components of the polarization tensor in the rest frame, $\mathcal{K}$ of a pulsar plasma are \citep{KML00,MGKF99}
\begin{align}
	 & \mT{\Pi}{1}{1} =\mT{\Pi}{2}{2}
	=-\sum_{\epsilon}\frac{e^2\tR{n}{\epsilon}}{m}\frac{1}{1+y^2}\left[\tr{\aV{\frac{1}{\gamma}}}{\epsilon}+\sum_{\alpha=\pm}\frac{\alpha(z-\beta_\alpha)^2\tr{R}{\epsilon}(\beta_\alpha)}{\beta_+-\beta_-}\right],
	\nn                                                                                                                                                                                                                                                            \\
	 & \mT{\Pi}{3}{3}=-\sum_{\epsilon}\frac{e^2\tR{n}{\epsilon}}{m}\left\{z^2\tr{W}{\epsilon}(z)+\frac{\tan^2\theta}{1+y^2}\left[\tr{\aV{\frac{1}{\gamma}}}{\epsilon}+\sum_{\alpha=\pm}\frac{\alpha\beta_\alpha^2\tr{R}{\epsilon}(\beta_\alpha)}{\beta_+-\beta_-}\right]
	\right\},
	\nn                                                                                                                                                                                                                                                            \\
	 & \mT{\Pi}{1}{3} =\mT{\Pi}{3}{1}
	=-\sum_{\epsilon}\frac{e^2\tR{n}{\epsilon}}{m}\frac{\tan\theta}{1+y^2}\left[-\tr{\aV{\frac{1}{\gamma}}}{\epsilon}
		+\sum_{\alpha=\pm}\frac{\alpha(z-\beta_\alpha)\beta_\alpha \tr{R}{\epsilon}(\beta_\alpha)}{\beta_+-\beta_-}
		\right],
	\nn                                                                                                                                                                                                                                                            \\
	 & \mT{\Pi}{1}{2} =-\mT{\Pi}{2}{1}
	=-\rmi\sum_{\epsilon}\frac{\epsilon e^2\tR{n}{\epsilon}}{m}\frac{y}{1+y^2}\sum_{\alpha=\pm}\frac{\alpha(z-\beta_\alpha)\tr{S}{\epsilon}(\beta_\alpha)}{\beta_+-\beta_-},
	\nn                                                                                                                                                                                                                                                            \\
	 & \mT{\Pi}{2}{3} =-\mT{\Pi}{3}{2}
	=\rmi\sum_{\epsilon}\frac{\epsilon e^2\tR{n}{\epsilon}}{m}\frac{y\tan\theta}{1+y^2}\sum_{\alpha=\pm}\frac{\alpha \beta_\alpha \tr{S}{\epsilon}(\beta_\alpha)}{\beta_+-\beta_-}.
	\label{Piij1}
\end{align}

The terms in Equation \eqref{eq:Kij1_and_NPP1} involving $\tr{R}{\epsilon}(\beta_\alpha)$ describe the contribution of the cyclotron resonances to non-gyrotropic dispersion and the terms involving $\tr{S}{\epsilon}(\beta_\alpha)$ describe the contribution of the cyclotron resonances to gyrotropic dispersion.

\section{Lorentz transformation between frames}
\label{sect:Lorentz}

In this section we discuss the Lorentz transformation between the rest frame $ \tr{\mathcal{K}}{\epsilon}$ of species $\epsilon$ and the pulsar frame $ \mathcal{K}'$, in which the species is streaming at speed $\tRr{\beta}{\epsilon}{\rm s}$ away from the star (positive direction).

\subsection{Lorentz transformation to the streaming frame}

Consider the Lorentz transformation between the rest frame $ \mathcal{K} $ of the plasma and the pulsar frame $ \mathcal{K}'$. We use 4-tensor notation with Greek indices $\mu$ running over (0,1,2,3) where $\mu=0$ denotes the time component and $\mu=i$ denotes the $i$th spatial component. An event is described by the (contravariant) 4-vector (in natural units with $c=1$) $x^\mu=[t,\bm{x}]$ in $ \mathcal{K}$ and $x^{\mu'}=[t',\bm{x}']$ in $ \mathcal{K}'$. The wave 4-vector is $k^\mu=[\omega,\bm{k}]$, with $ \bm{k} = (k_\perp, 0, k_\parallel) $, in $ \mathcal{K}$ and $k^{\mu'}=[\omega',\bm{k}']$ in $ \mathcal{K}'$. The covariant components are $x_\mu=[t,-\bm{x}]$ and $k_\mu=[\omega,-\bm{k}]$.

The Lorentz transformation matrices between the frames are
\begin{equation}
	L^{\mu'}{}_\mu(-\beta_{\rm s})=
	\left(\begin{array}{cccc}
			\gamma_{\rm s}               & 0 & 0 & \gamma_{\rm s} \beta_{\rm s} \\
			0                            & 1 & 0 & 0                            \\
			0                            & 0 & 1 & 0                            \\
			\gamma_{\rm s} \beta_{\rm s} & 0 & 0 & \gamma_{\rm s}
		\end{array}
	\right),
	\qquad
	L^\mu{}_{\mu'}(-\beta_{\rm s})=
	\left(\begin{array}{cccc}\gamma_{\rm s}          & 0 & 0 & -\gamma_{\rm s} \beta_{\rm s} \\
             0                             & 1 & 0 & 0                             \\
             0                             & 0 & 1 & 0                             \\
             -\gamma_{\rm s} \beta_{\rm s} & 0 & 0 & \gamma_{\rm s}
		\end{array}
	\right),
	\label{1a.16}
\end{equation}
with $\gamma_{\rm s}=(1-\beta_{\rm s}^2)^{-1/2}$.
The non-zero components of $L^{\mu'}{}_\mu=L^{\mu'}{}_\mu(-\beta_{\rm s})$ and $L^\mu{}_{\mu'}=L^\mu{}_{\mu'}(-\beta_{\rm s})$ are
\bea
L^{0'}{}_{0}&&=L^{3'}{}_{3}=L^{0}{}_{0'}=L^{3}{}_{3'}=\gamma_{\rm s},\qquad
L^{1'}{}_{1}=L^{2'}{}_{2}=L^{1}{}_{1'}=L^{2}{}_{2'}=1,
\nn\\
L^{0'}{}_{3}&&=L^{3'}{}_{0}=\gamma_{\rm s}\beta_{\rm s}, \hspace{26.5mm} L^{0}{}_{3'}=L^{3}{}_{0'}=-\gamma_{\rm s}\beta_{\rm s}.
\qquad
\label{Lmunup}
\eea

The frequency $\omega$ and the components $k_\parallel$ and $k_\perp$, parallel and perpendicular, respectively, to the magnetic field transform to
\begin{equation}
	\omega'=\gamma_{\rm s}(\omega+k_\parallel c\beta_{\rm s}),
	\quad
	k'_\parallel c=\gamma_{\rm s}(k_\parallel c+\omega\beta_{\rm s}),
	\quad
	k'_\perp=k_\perp.
	\label{LT1a}
\end{equation}
In terms of the variables $z=\omega/k_\parallel c$ and $\theta=\arctan(k_\perp/k_\parallel)$ in the unprimed frame, $ \mathcal{K} $, and $z'=\omega'/k'_\parallel c$ and $\theta'=\arctan(k'_\perp/k'_\parallel)$ in the primed frame, $ \mathcal{K}' $, equations \eqref{LT1a} and the inverse transforms imply
\begin{equation}
	z'=\frac{z+\beta_{\rm s}}{1+\beta_{\rm s} z},
	\quad
	z=\frac{z'-\beta_{\rm s}}{1-\beta_{\rm s} z'},
	\quad
	\tan\theta'=\frac{\tan\theta}{\gamma_{\rm s}(1+\beta_{\rm s} z)},
	\quad
	\tan\theta=\frac{\tan\theta'}{\gamma_{\rm s}(1-\beta_{\rm s} z')}.
	\label{LT2a}
\end{equation}

\subsection{Lorentz transformation of the response tensor}
\label{sect:LTKij}

Several steps are involved in Lorentz transforming the dielectric tensor  \citep[e.g.,][]{1973PlPh...15...99M} as discussed by \cite{RMM2} in the case where the cyclotron resonances are neglected.

The first step is to write the dielectric tensor in the form \eqref{eq:Kij1_and_NPP1} and to note that the 3-tensor $\Pi_{ij}$ may be interpreted as the space components of the 4-tensor $\Pi^{\mu}{}_{\nu}(k)$ that relates the 4-current $J^\mu(k)$, to 4-potential, $A^\nu(k)$; specifically, the space components of $J^\mu(k)=\Pi^\mu{}_\nu(k)A^\nu(k)$ imply the relation between the 3-current and the vector potential (in the temporal gauge), $J^i(k)=\Pi^i{}_j(k)A^j(k)$, where the argument $k$ denotes the components of $k^\mu$. The 3-tensor $\Pi_{ij}$ is term-by-term equal to the mixed components of the 4-tensor $\Pi^{\mu}{}_{\nu}$ with $\mu=i$, $\nu=j$.\footnote{Note that in 4-tensor notation it is conventional to distinguish components in the primed frame, in which the plasma is streaming, by primes on the indices, such that $\Pi^{\mu'}{}_{\nu'}$ are the transformed mixed tensor components in $\mathcal{K}'$. Although the corresponding space components are $\Pi^{i'}{}_{j'}$, in 3-tensor notation it is conventional to denote the transformed components by a prime on the kernel symbol, so that $\Pi^{i'}{}_{j'}$ is term-by-term equal to the 3-tensor $\Pi'_{ij}$.}

The next step is to construct the full 4-tensor $\Pi^{\mu}{}_{\nu}$ from the space components $\Pi^{i}{}_{j}$ using the charge-continuity and gauge-invariance relations, $k_\mu\Pi^\mu{}_\nu=0$ and $k^\nu\Pi^\mu{}_\nu=0$, respectively. The 4-tensor components $\Pi^0{}_0$, $\Pi^i{}_0$, $\Pi^0{}_j$ are given in terms of the mixed tensor components by
\begin{equation}
	\Pi^\mu{}_0=-\frac{\tan\theta}{z}\Pi^\mu{}_1-\frac{1}{z}\Pi^\mu{}_3, \qquad
	\Pi^0{}_\nu=\frac{\tan\theta}{z}\Pi^1{}_\nu+\frac{1}{z}\Pi^3{}_\nu.
	\label{Pi0s}
\end{equation}
The third step is to apply the Lorentz transformation to $\Pi^{\mu}{}_{\nu}(k)$ in $ \mathcal{K}$, to find $\Pi^{\mu'}{}_{\nu'}(k')=L^{\mu'}{}_{\mu}\Pi^{\mu}{}_{\nu}(L^{-1}[k])L^{\nu}{}_{\nu'}$ in $ \mathcal{K}'$, where $k'=L^{-1}[k]$ denotes the components $\omega',\bm{k}'$ expressed in terms of $\omega,\bm{k}$. The transformed components $\Pi^{i'}{}_{j'}$ are then identified (term-by-term) as the components of the (polarization) 3-tensor $\Pi'_{ij}$ in $ \mathcal{K}'$. Finally, the dielectric tensor in $ \mathcal{K}'$ is identified as
\begin{equation}
	K'_{ij}(\omega',\bm{k}')=\delta_{ij}+\frac{\Pi'_{ij}(\omega',\bm{k}')}{\varepsilon_0\omega'^2},
	\label{Kijp}
\end{equation}
which is interpreted as the dielectric tensor for the streaming distribution in the primed frame.

\subsection{Transformed polarization tensor}

The Lorentz transformation applied to the polarization 3-tensor $\Pi_{ij}$ gives the transformed 3-tensor $\Pi'_{ij}$:
\begin{equation}
	\begin{gathered}
		\Pi'_{11}
		= \Pi_{11},\quad
		\Pi'_{12}
		= \Pi_{12},\quad
		\Pi'_{21}
		= \Pi_{21},\quad
		\Pi'_{22}
		= \Pi_{22},\\
		\Pi'_{13}
		= \gamma_s \left[\frac{\beta_s \tan\theta}{z} \Pi_{11} + \frac{z + \beta_s}{z} \Pi_{13}\right],\quad
		\Pi'_{31}
		= \gamma_s \left[\frac{\beta_s \tan\theta}{z} \Pi_{11} + \frac{z + \beta_s}{z} \Pi_{31}\right],\\
		\Pi'_{23}
		= \gamma_s \left[\frac{\beta_s \tan\theta}{z} \Pi_{21} + \frac{z + \beta_s}{z} \Pi_{23}\right],\quad
		\Pi'_{32}
		= \gamma_s \left[\frac{\beta_s \tan\theta}{z} \Pi_{12} + \frac{z + \beta_s}{z} \Pi_{32}\right],\\
		\Pi'_{33}
		= \gamma_s^2 \left[\left(\frac{\beta_s \tan\theta}{z}\right)^2 \Pi_{11}
  + \left(\frac{z + \beta_s}{z}\right) \left(\frac{\beta_s \tan\theta}{z}\right) \left(\Pi_{13} + \Pi_{31}\right) + \left(\frac{z + \beta_s}{z}\right)^2 \Pi_{33}\right].
	\end{gathered}
	\label{LT4}
\end{equation}
With $\Pi_{ij}$ given by equation \eqref{Piij1}, the transformed tensor \eqref{LT4} becomes
\begin{align}
	 & {\Pi'}_{11} ={\Pi'}_{22}
	=-\sum_{\epsilon}\frac{e^2\tR{n}{\epsilon}}{m}\frac{1}{1+y^2}\left[\tr{\aV{\frac{1}{\gamma}}}{\epsilon}+\sum_{\alpha=\pm}\frac{\alpha(z-\beta_\alpha)^2\tr{R}{\epsilon}(\beta_\alpha)}{\beta_+-\beta_-}\right],
	\nn                                                                                                                                                                                                                                                                                                         \\
	 & {\Pi'}_{33}=-\gamma_{\rm s}^2\sum_{\epsilon}\frac{e^2\tR{n}{\epsilon}}{m}\bigg\{(z+\beta_{\rm s})^2\tr{W}{\epsilon}(z)
  \nn
  \\
  &\qquad\qquad\qquad\qquad\qquad
  +\frac{\tan^2\theta}{1+y^2}\left[\tr{\aV{\frac{1}{\gamma}}}{\epsilon}+\sum_{\alpha=\pm}\frac{\alpha(\beta_{\rm s}+\beta_\alpha)^2\tr{R}{\epsilon}(\beta_\alpha)}{\beta_+-\beta_-}\right]
	\bigg\},
	\nn                                                                                                                                                                                                                                                                                                         \\
	 & {\Pi'}_{13} ={\Pi'}_{31}
	=-\gamma_{\rm s}\sum_{\epsilon}\frac{e^2\tR{n}{\epsilon}}{m}\frac{\tan\theta}{1+y^2}\left[-\tr{\aV{\frac{1}{\gamma}}}{\epsilon}
		+\sum_{\alpha=\pm}\frac{\alpha(z-\beta_\alpha)(\beta_{\rm s}+\beta_\alpha)\tr{R}{\epsilon}(\beta_\alpha)}{\beta_+-\beta_-}
		\right],
	\nn                                                                                                                                                                                                                                                                                                         \\
	 & {\Pi'}_{12} =-{\Pi'}_{21}
	=-\rmi\sum_{\epsilon}\frac{\epsilon e^2\tR{n}{\epsilon}}{m}\frac{y}{1+y^2}\sum_{\alpha=\pm}\frac{\alpha(z-\beta_\alpha)\tr{S}{\epsilon}(\beta_\alpha)}{\beta_+-\beta_-},
	\nn                                                                                                                                                                                                                                                                                                         \\
	 & {\Pi'}_{23} =-{\Pi'}_{32}
	=\rmi\gamma_{\rm s}\sum_{\epsilon}\frac{\epsilon e^2\tR{n}{\epsilon}}{m}\frac{y\tan\theta}{1+y^2}\sum_{\alpha=\pm}\frac{\alpha(\beta_{\rm s}+\beta_\alpha)\tr{S}{\epsilon}(\beta_\alpha)}{\beta_+-\beta_-}.
	\label{Piij1a}
\end{align}
The result \eqref{Piij1a} is the polarization tensor in the frame $\mathcal{K}'$, expressed in terms of the plasma parameters in the (unprimed) rest frame $\mathcal{K}$ of the streaming distribution. An alternative form for this tensor is obtained by re-expressing \eqref{Piij1a} in terms of plasma parameters and RPDFs in the primed frame.

\subsection{Lorentz-transformed RPDFs}

In order to rewite \eqref{Piij1a} in terms of variables in $\mathcal{K}'$, one need to relate the three RPDFs $\tr{W}{\epsilon}(z)$, $\tr{R}{\epsilon}(\beta_\alpha)$, $\tr{S}{\epsilon}(\beta_\alpha)$ as defined in $ \mathcal{K}$ to the corresponding RPDFs in $ \mathcal{K}'$. The latter are defined as
\begin{equation}\label{WRS2}
	\tRr{W}{\prime}{\epsilon}(z')
	= \tRr{\aV{\frac{1}{\gamma'^3(\beta'-z')^2}}}{\prime}{\epsilon},\quad
	\tRr{R}{\prime}{\epsilon}(\beta_\alpha')
	= \tRr{\aV{\frac{1}{\gamma'(\beta'-\beta_\alpha')}}}{\prime}{\epsilon},\quad
	\tRr{S}{\prime}{\epsilon}(\beta_\alpha')
	= \tRr{\aV{\frac{1}{\gamma'^2(\beta'-\beta_\alpha')}}}{\prime}{\epsilon}.
\end{equation}
The averages of any quantity $X$ in the two frames are defined by \eqref{eq:average},  which implies
\begin{equation}
	\tRr{\aV{\frac{X}{\gamma'}}}{\prime}{\epsilon} = \frac{1}{\gamma_{\rm s}} \tr{\aV{\frac{X}{\gamma}}}{\epsilon}.
	\label{avQ/g}
\end{equation}
Using \eqref{avQ/g} and the relations \eqref{LT2a}, one finds
\begin{equation}\label{WRS3}
	\begin{split}
		\tRr{W}{\prime}{\epsilon}(z')
		& = \gamma_{\rm s} (1 + \beta_{\rm s} z)^2 \tr{W}{\epsilon}(z),\\
		\tRr{R}{\prime}{\epsilon}(\beta'_\alpha)
		& = \gamma_{\rm s} (1 + \beta_{\rm s} \beta_\alpha) \left[(1 + \beta_{\rm s} \beta_\alpha) \tr{R}{\epsilon}(\beta_\alpha) + \beta_{\rm s} \tr{\aV{\frac{1}{\gamma}}}{\epsilon}\right],\\
		\tRr{S}{\prime}{\epsilon}(\beta'_\alpha)
		& = (1 + \beta_{\rm s} \beta_\alpha) \tr{S}{\epsilon}(\beta_\alpha),
	\end{split}
\end{equation}
and the inverse transforms
\begin{equation}\label{WRS4}
	\begin{split}
		\tr{W}{\epsilon}(z)
		& = \gamma_{\rm s}^3(1 - \beta_{\rm s} z')^2\tRr{W}{\prime}{\epsilon}(z'),\\
		\tr{R}{\epsilon}(\beta_\alpha)
		& = \gamma_{\rm s}^3(1 - \beta_{\rm s} \beta'_\alpha)\left[(1 - \beta_{\rm s} \beta'_\alpha)\tRr{R}{\prime}{\epsilon}(\beta'_\alpha) - \beta_{\rm s} \tRr{\aV{\frac{1}{\gamma'}}}{\prime}{\epsilon}\right],\\
		\tr{S}{\epsilon}(\beta_\alpha)
		& = \gamma_{\rm s}^2(1 - \beta_{\rm s} \beta'_\alpha)\tRr{S}{\prime}{\epsilon}(\beta'_\alpha).
	\end{split}
\end{equation}
The identity \eqref{avQ/g} implies $\gamma_{\rm s} \tRr{\aV{1/\gamma'}}{\prime}{\epsilon}=\tr{\aV{1/\gamma}}{\epsilon}$.

\subsection{The polarization tensor in the primed frame}

The response tensor $\Pi'_{ij}(\omega',\bm{k}')$ in $\mathcal{K}'$ is only partly determined by the equalities \eqref{Piij1}; one also needs to express the unprimed parameters in terms of the primed parameters. In addition to the identities \eqref{LT1a}, \eqref{LT2a}, \eqref{LT3}, and \eqref{WRS4} the following identities are useful in this context:
\begin{equation} \label{identities}
\begin{gathered}
    \gamma_{\rm s}(1-z'\beta_{\rm s})
        = \frac{1}{\gamma_{\rm s}(1+z\beta_{\rm s})},
    \quad
    y
        = \frac{y'}{\gamma_{\rm s}(1-z'\beta_{\rm s})},
    \quad
    1+y'^2-z'^2
        = \frac{1+y^2-z^2}{\gamma_{\rm s}^2 (1+z\beta_{\rm s})^2},
    \\
    \beta'_\pm
        = \frac{z'\pm y'[1+y'^2-z'^2]^{1/2}}{1+y'^2}=\frac{\beta_\pm+\beta_{\rm s}}{1+\beta_\pm\beta_{\rm s}},
    \quad
    z-\beta_\pm
        = \frac{z'-\beta'_\pm}{\gamma_{\rm s}^2(1-z'\beta_{\rm s})(1-\beta'_\pm\beta_{\rm s})},
    \\
    \beta_+-\beta_-
        = \frac{\beta'_+-\beta'_-}{\gamma_{\rm s}^2(1+\beta_+\beta_{\rm s})(1+\beta_-\beta_{\rm s})},
    \quad
    (1+y^2)(\beta_+-\beta_-)
        = \frac{(1+y'^2)(\beta'_+-\beta'_-)}{\gamma_{\rm s}^2(1-z'\beta_{\rm s})^2},
    \\
    \beta_+\beta_-
        = \frac{z^2-y^2}{1+y^2},
    \quad
    \frac{1}{(1-\beta'_+\beta_{\rm s})(1-\beta'_-\beta_{\rm s})}
        = \frac{\gamma_{\rm s}^2(1-z'\beta_{\rm s})^2+y'^2}{\gamma_{\rm s}^2(1+y'^2)}.
\end{gathered}
\end{equation}

Using these expressions, a lengthy calculation leads to the following expression for $\Pi'_{ij}(\omega',\bm{k}')$:
\bea
\Pi'_{11}&=&-\sum_{\epsilon}\frac{e^2\tR{n}{\epsilon \prime}}{\gamma_{\rm s}m(1+y'^2)}\bigg[
	\tr{\aV{\frac{1}{\gamma}}}{\epsilon}
	+\sum_{\alpha=\pm}\frac{\alpha(z'-\beta'_\alpha)^2\tRr{R}{\prime}{\epsilon}(\beta'_\alpha)}{\beta'_+-\beta'_-}
	\bigg],
\nn\\
\Pi'_{12}
&=&-\rmi\sum_{\epsilon}\frac{\epsilon e^2\tR{n}{\epsilon \prime}}{\gamma_{\rm s}m}\frac{y'}{1+y'^2}\sum_{\alpha=\pm}\frac{\alpha(z'-\beta'_\alpha)\tRr{S}{\prime}{\epsilon}(\beta'_\alpha)}{\beta'_+-\beta'_-},
\nn\\
\Pi'_{23}
&=&\rmi\sum_{\epsilon}\frac{\epsilon e^2\tR{n}{\epsilon \prime}}{\gamma_{\rm s}m}\frac{y'\tan\theta'}{1+y'^2}\sum_{\alpha=\pm}\frac{\alpha \beta'_\alpha \tRr{S}{\prime}{\epsilon}(\beta'_\alpha)}{\beta'_+-\beta'_-},
\nn\\
\Pi'_{13}&=&-\sum_{\epsilon}\frac{e^2\tR{n}{\epsilon \prime}\tan\theta'}{\gamma_{\rm s}m(1+y'^2)}
\bigg[-\tr{\aV{\frac{1}{\gamma}}}{\epsilon}
	+\sum_{\alpha=\pm}\frac{\alpha(z'-\beta'_\alpha)\beta'_\alpha \tRr{R}{\prime}{\epsilon}(\beta'_\alpha)}{\beta'_+-\beta'_-}
	\bigg],
\nn\\
\Pi'_{33}&=&\sum_{\epsilon}\frac{e^2\tR{n}{\epsilon \prime}}{\gamma_{\rm s}m}\left\{z'^2\tRr{W}{\prime}{\epsilon}(z')
+\frac{\tan^2\theta'}{1+y'^2}\left[
	\tr{\aV{\frac{1}{\gamma}}}{\epsilon}
	+\sum_{\alpha=\pm}\frac{\alpha\beta'^2_\alpha \tRr{R}{\prime}{\epsilon}(\beta'_\alpha)}{\beta'_+-\beta'_-}\right]\right\}.
\qquad
\label{Pips}
\eea
where the substitution $\tr{\aV{1/\gamma}}{\epsilon}=\gamma_{\rm s}\tRr{\aV{1/\gamma'}}{\prime}{\epsilon}$ is not made explicitly.

\section{Alternative evaluation of the response tensor in the primed frame}
An alternative method of including the streaming involves evaluating the response tensor directly in the primed frame. This involves Lorentz transforming both the distribution function and and the tensor $A_{ij}(\omega, \bm{k};\beta)$ to the primed frame.

\subsection{Lorentz transforming the distribution function}
Any distribution function, $\tr{f}{\epsilon}(\bm{p})$ is a Lorentz invariant. In the 1D case, the distribution function in $ \mathcal{K}$ may be written as $\tr{g}{\epsilon}(u)$, with $u^\mu=p^\mu/m=(\gamma,u\bm{b})$, where $\bm{b}$ is the unit vector along the magnetic field and $u=\gamma\beta$ is the 4-speed. The Lorentz transformation implies
\begin{equation}
	\gamma'=\gamma\gamma_{\rm s}(1+\beta\beta_{\rm s}),
	\quad
	\beta'=\frac{\beta+\beta_{\rm s}}{1+\beta\beta_{\rm s}};
	\qquad
	\gamma=\gamma'\gamma_{\rm s}(1-\beta'\beta_{\rm s}),
	\quad
	\beta=\frac{\beta'-\beta_{\rm s}}{1-\beta'\beta_{\rm s}}.
	\label{LT3}
\end{equation}

The distribution function in $ \mathcal{K}'$ may be written as $\tRr{g}{ \prime}{\epsilon}(u')$, with $u'=\gamma'\beta'$. The normalizations in the two frames are
\begin{equation}
	\int_{-\infty}^\infty \rmd u\,\tr{g}{\epsilon}(u)=\tR{n}{\epsilon},
	\qquad
	\int_{-\infty}^\infty \rmd u'\,\tRr{g}{\prime}{\epsilon}(u')=\tR{n}{\epsilon}^{\prime}.
	\label{nnp}
\end{equation}

A 1D J\"uttner distribution is $\tr{g}{\epsilon}(u)=\tR{n}{\epsilon} \exp(-\rho\gamma)/2K_1(\rho)$ where $\rho=mc^2/T$ is the inverse temperature in units of the electron rest energy. Transforming to $ \mathcal{K}'$ gives
\begin{equation}
	\tRr{g}{\prime}{\epsilon}(u')
	= \frac{\tR{n}{\epsilon \prime}}{\gamma_{\rm s}} \frac{e^{-\rho\gamma_{\rm s}\gamma'(1-\beta_{\rm s}\beta')}}{2K_1(\rho)},
	\qquad
	\tR{n}{\epsilon \prime}={\gamma_{\rm s}}\tR{n}{\epsilon}.
	\label{Juttnerp}
\end{equation}
This result follows, for $\tr{g}{\epsilon}(-u)=\tr{g}{\epsilon}(u)$, from $du'=d(\beta'\gamma')=\gamma'^3d\beta'$, $du=\gamma^3d\beta$ and $d\beta'/d\beta=\gamma^2/\gamma'^2$ implying $du'/du=\gamma'/\gamma$, with $\gamma'$ and $\beta'$ given in terms of $\gamma$ and $\beta$ by equation \eqref{LT3}. The distribution function \eqref{Juttnerp} may be interpreted as a streaming J\"uttner distribution function in the primed frame.

\subsubsection{Response tensor for streaming J\"uttner distribution}

The response tensor for the streaming J\"uttner distribution function \eqref{Juttnerp}, evaluated in the frame $\mathcal{K}'$ is
\begin{equation}
	\Pi'_{ij}(\omega,\bm{k})=-\sum_{\epsilon}\frac{e^2}{m}\int_{-\infty}^\infty du'\,\tRr{g}{\prime}{\epsilon}(u')
	\frac{A'_{ij}(\omega',\bm{k}';\beta')}{\gamma'}.
	\label{NPP1s}
\end{equation}
where $A'_{ij}/\gamma'$ is related to $A_{ij}/\gamma$ by the Lorentz transformation that relates $\Pi'_{ij}$ to $\Pi_{ij}$, cf.\ equation \eqref{LT4}.

The integral in equation \eqref{NPP1s} may be reduced to the same form as the integral in the non-streaming case by changing the variable of integration from $u$ to $u'$, with $du'/du=\gamma'/\gamma$ and replacing $du'\,\tRr{g}{\prime}{\epsilon}(u')\gamma'$ by $du\tr{g}{\epsilon}(u)\gamma$, where $\tr{g}{\epsilon}(u)$ is the non-streaming distribution function. The resulting expression for the response tensor reproduces the form \eqref{Pips}.

The primed frame, $\mathcal{K}'$, may be interpreted as the pulsar frame, provided that the electrons and positrons stream at the same speed (and that the rotation of the pulsar plasma is neglected). The transformed polarization tensor, in either form \eqref{Piij1a} or \eqref{Pips}, may then be interpreted as the polarization tensor in the pulsar frame, expressed in terms of the unprimed and primed variables, respectively. The form \eqref{Pips} in terms of the primed variables might appear to be the more convenient because all quantities are defined in the frame of relevance to the observer. However, we find the form \eqref{Piij1a} to be more convenient because the RPDFs are usually defined in the rest frame of the distribution of particles. Alternatively one may use a mixed notation: starting from the form \eqref{Pips} one may use the relations \eqref{WRS3} to rewrite the RPDFs in the primed frame in terms of the RPDFs in the unprimed frame, with the unprimed variables, including the arguments of the RPDFs, rewritten in terms of the primed variables using \eqref{LT3}.

\section{Approximate and limiting cases}
Limiting cases of the response tensor include limits related to the distribution function, and limits related to the wave dispersion. Two limits of the J\"uttner distribution are the cold-plasma case, $\rho\gg1$, and the high-energy  limit, $\rho\ll1$.

\subsection{Cold-plasma limit}
The cold-plasma limits of the RPDFs follow from the definitions \eqref{WRS1}, with $ du\, \tr{g}{\epsilon}(u) \to du\, \tR{n}{\epsilon}\, \delta(u) $. The limiting forms are
\begin{equation}
	\tr{W}{\epsilon}(\tR{z}{\epsilon})
        \to \frac{1}{\tR{z}{\epsilon 2}},
	\qquad
	\tr{R}{\epsilon}(\tRr{\beta}{\epsilon}{\alpha})
        \to -\frac{1}{\tRr{\beta}{\epsilon}{\alpha }},
	\qquad
	\tr{S}{\epsilon}(\tRr{\beta}{\epsilon}{\alpha}) 
        \to -\frac{1}{\tRr{\beta}{\epsilon}{\alpha }}.
	\label{WRSc}
\end{equation}
For a cold streaming plasma, these are replaced by
\begin{equation}
	\tRr{W}{\prime}{\epsilon}(\tR{z}{\epsilon \prime})
        \to \frac{1}{\tRr{\gamma}{\epsilon 3}{\rm s} (\tR{z}{\epsilon \prime} - \tRr{\beta}{\epsilon}{\rm s})^2},
	\quad
	\tRr{R}{\prime}{\epsilon}(\tRr{\beta}{\epsilon \prime}{\alpha})
        \to - \frac{1}{\tRr{\gamma}{\epsilon}{\rm s} (\tRr{\beta}{\epsilon \prime}{\alpha} - \tRr{\beta}{\epsilon}{\rm s})},
	\quad
	\tRr{S}{\prime}{\epsilon}(\tRr{\beta}{\epsilon \prime}{\alpha})
        \to - \frac{1}{\tRr{\gamma}{\epsilon 2}{\rm s} (\tRr{\beta}{\epsilon \prime}{\alpha} - \tRr{\beta}{\epsilon}{\rm s})},
	\label{WRScs}
\end{equation}
using the limit $ du'\, \tRr{g}{\prime}{\epsilon}(u') \to du'\, \tRr{n}{\epsilon \prime}{}\, \delta(u' - \tRr{u}{\epsilon \prime}{}) $.

The cold-plasma limit of the polarization tensor \eqref{eq:Kij1_and_NPP1} for a non-streaming J\"uttner distribution reduces to
\begin{equation}
	\Pi_{ij}(\omega,\bm{k})
        \to - \sum_{\epsilon} \frac{e^2\tR{n}{\epsilon}}{m}A_{ij}(\omega,\bm{k};0),
	\label{Piijc}
\end{equation}
with $A_{ij}(\omega,\bm{k};\beta)$ given by \eqref{eq:Aij1}. The result \eqref{Piijc} is reproduced by setting $g(u) \to \tR{n}{\epsilon} \delta(u)$ in \eqref{eq:Kij1_and_NPP1}.

The cold-plasma limit for a streaming J\"uttner distribution reduces to
\begin{equation}
	\Pi'_{ij}(\omega,\bm{k})
        \to \sum_{\epsilon} \frac{e^2\tR{n}{\epsilon}}{\gamma_{\rm s}m}A_{ij}(\omega',\bm{k}'; \beta_{\rm s}),
	\label{Piijcs}
\end{equation}
with $\omega'=\gamma_{\rm s}(\omega-k_\parallel \beta_{\rm s}c)$, $k_\perp^{\prime}=k_\perp$, $k_\parallel'=\gamma_{\rm s}(k_\parallel-\omega\beta_{\rm s}/c)$. The result \eqref{Piijcs}) is implied directly by setting $g(u) \to \tR{n}{\epsilon} \delta(u-u_{\rm s})$ in \eqref{eq:Kij1_and_NPP1}.

\subsection{The highly relativistic limit $\rho\ll1$}
In the highly relativistic limit, $\rho\ll1$ the RPDFs by be approximated by exponential-integral functions. One finds
\citep{Luo_Melrose_2004_2004JPlPh..70..709L,2004PhRvE..70a6404M}
\bea
&&W(z,\rho)=-\rho\gamma_\phi^2\left\{1-\half\rho\gamma_\phi z^2
\left[e^{-\rho\gamma_\phi}{\rm Ei}(\rho\gamma_\phi)-e^{\rho\gamma_\phi}{\rm Ei}(-\rho\gamma_\phi)
\right]\right\}.
\nn\\
&&R(z,\rho)=-\half\rho\gamma_\phi^2z
\left[e^{-\rho\gamma_\phi}{\rm Ei}(\rho\gamma_\phi)+e^{\rho\gamma_\phi}{\rm Ei}(-\rho\gamma_\phi)
\right],
\nn\\
&&S(z,\rho)=-\half\rho\gamma_\phi z
\left[e^{-\rho\gamma_\phi}{\rm Ei}(\rho\gamma_\phi)-e^{\rho\gamma_\phi}{\rm Ei}(-\rho\gamma_\phi)
\right].
\qquad
\label{RSWei}
\eea
For $\rho\gamma_\phi=\rho(1-z^2)^{-1/2}\gg1$ one has
\begin{align}
    W(z,\rho)
        & \sim -\rho \gamma_\phi^2 \left\{1 - z^2 \sum_{k = 0}^{\infty} \frac{(2k)!}{(\rho \gamma_\phi)^{2k}}\right\}
        \approx -\rho \gamma_\phi^2 \left\{1 - z^2 \left[1 + \frac{2}{(\rho\gamma_\phi)^2}\right]\right\},\\
    R(z,\rho)
        & \sim -\frac{z}{\rho} \sum_{k = 0}^{\infty} \frac{(2k+1)!}{(\rho \gamma_\phi)^{2k}}
        \approx -\frac{z}{\rho}\left[1+\frac{6}{(\rho\gamma_\phi)^2}\right],\\
    S(z,\rho)
        & \sim - z \sum_{k = 0}^{\infty} \frac{(2k)!}{(\rho \gamma_\phi)^{2k}}
        \approx -z\left[1 + \frac{2}{(\rho\gamma_\phi)^2}\right]
        \label{rhoggg1}
\end{align}

\subsection{Approximations to the arguments of RPDFs}
\label{approxarg}

The arguments of the RPDFs $R$ and $S$ in \eqref{Piij1a} may be approximated in two ways: by making the WAA and by assuming highly relativistic streaming, $\gamma_{\rm s}\gg1$. The first of these approximations follows from $z'\approx1/\cos\theta'$ and $y'\approx1/x'\cos\theta'$, giving
\begin{equation}\label{betapmsapprox}
    \beta'_\alpha
        \approx \frac{x^{\prime 2} \cos\theta' + \alpha \left[1 - x^{\prime 2} \sin^2\theta'\right]^{1/2}}{1 + x^{\prime 2} \cos^2\theta'},\quad
    \gamma'_\alpha
        \approx s \frac{1 + \alpha \abs{\cos\theta'} \left[1 - x^{\prime 2} \sin^2\theta'\right]^{1/2}}{x' \sin^2\theta'},
\end{equation}
where $x'=\omega'/\Omega_e$ is the ratio of the wave frequency to the cyclotron frequency in $\mathcal{K}'$. The second approximation follows by assuming $\beta_{\rm s}\approx 1-1/2\gamma_{\rm s}^2$, giving
\begin{equation}
	\frac{\beta'_\alpha-\beta_{\rm s}}{1-\beta'_\alpha\beta_{\rm s}}
	\approx-1+\frac{1}{2\Gamma_\alpha^2},
	\qquad
	\Gamma_\alpha^2=2\gamma_{\rm s}^2\frac{1-\beta'_\alpha}{1+\beta'_\alpha},
	\label{betapmsapprox}
\end{equation}
where $\gamma_{\rm s}^2\gg(\gamma'_\alpha)^2$ is assumed. These approximations may be applied to the exponential integral approximation \eqref{rhoggg1} to the RPDFs
with $z\to(\beta'_\alpha-\beta_{\rm s})/(1-\beta'_\alpha\beta_{\rm s})$ and $\gamma_\phi\to\Gamma_\alpha$.

\section{Discussion and Conclusions}

Our primary purpose in this article is to derive the response tensor for a relativistically streaming J\"uttner distribution, which we argue is the preferred model for a pulsar plasma. The expression \eqref{Pips} for the polarization tensor, $\Pi'_{ij}$, is the desired result. The corresponding dielectric tensor, $K'_{ij}$, follows by inserting this expression into equation \eqref{eq:Kij1_and_NPP1}, modified by adding superscripts ${\rm s}$ to $K_{ij}$ and $\Pi_{ij}$. Using the general form \eqref{Pips} for the response tensor allows a general treatment of wave dispersion in a pulsar plasma that includes the relativistic streaming, the intrinsic spread in Lorentz factors in the plasma rest frame and the cyclotron resonances. We plan to discuss the details of the wave dispersion elsewhere.  

The response tensor \eqref{Pips} involves three RPDFs, defined in the rest frame of the J\"uttner distribution. Approximations to the response tensor involve relevant approximations to these RPDFs. We note two opposite limits of the three RPDFs: the well-known cold-plasma limit \eqref{WRSc}, and the highly relativistic limit \eqref{RSWei}. Another type of approximation involves the arguments of the RPDFs $R,S$ that describe the effect of the cyclotron resonances. In the absence of streaming these arguments are $\beta_\pm$, given by \eqref{eq:beta_pm} and plotted in Figure~\ref{fig:beta_pm_contours_yz}. The arguments of $R,S$ when the streaming is included are Lorentz transformed to $\beta'_\pm$, given by \eqref{betazpm}. For highly relativistic streaming $\beta'_\pm$ are highly relativistic $\gamma'_\pm\gg1$ except for a tiny range of parameters corresponding $\beta_\pm\approx\beta_{\rm s}$. Then $R,S$ may be approximated by their highly relativistic limits, as discussed in \S\ref{approxarg}, leading to substantial simplification.

A particular motivation for the theory developed here is an application to generalized Faraday rotation (GFR) in pulsars and magnetars. Such GFR occurs in a region where the frequency of the escaping radiation is much greater that the plasma and cyclotron frequencies. and the wave dispersion may then be treated in the weak-anisotropy approximation (WAA), in which the refractive indices are assumed close to unity and the polarization close to transverse. In the WAA, both the refractive indices and polarizations of the two natural wave modes, and hence the GFR axis and the rate of rotation per unit length (of the ray path) of the polarization point about it, are determined to first order in this expansion. Based on the general form of the response tensor derived here, we discuss GFR in the WAA in a separate paper.


\bibliography{CEAPReferences}
\bibliographystyle{aasjournal}

\newpage
\appendix

\section{Different streaming speeds, $\tRr{\beta}{\epsilon}{\rm s}$}
Elliptical polarization in a pulsar plasma can be due to either a net charge density, $\eta$, or a net current density, $\bm{J}$. In order to include the effect of the pulsar current on the wave dispersion, we generalize the foregoing model to allow for different streaming speeds for the electrons and positrons, $\beta_{\rm s}\to\tRr{\beta}{\epsilon}{\rm s}$ say, with $\tRr{\beta}{+}{\rm s} \ne \tRr{\beta}{-}{\rm s} $. Then one has
\begin{equation}
	\eta=\sum_\epsilon \epsilon e\tR{n}{\epsilon},
	\qquad
	\bm{J}=\sum_\epsilon \epsilon ec\tR{n}{\epsilon}\tRr{\beta}{\epsilon}{\rm s}\bm{b}.
	\label{etaJ}
\end{equation}
Writing
\begin{equation}
	\tR{n}{\epsilon}={\bar n}+\half\epsilon\delta n,
	\qquad
	\tRr{\beta}{\epsilon}{\rm s}={\bar\beta}_{\rm s}+\half\epsilon\delta\beta_{\rm s},
	\label{diffs}
\end{equation}
equations \eqref{etaJ} become
\begin{equation}
	\eta=e\delta n,
	\qquad
	\bm{J}= ec(\bar{n} \delta\beta_{\rm s} + \delta n \bar{\beta}_{\rm s})\bm{b}.
	\label{etaJ1}
\end{equation}

We introduce two rest frames, $\tr{\mathcal{K}}{\epsilon}$, one for each of the two species $\epsilon=\pm$. To avoid confusion we continue to denote the pulsar frame by $\mathcal{K}'$. The response tensor (dielectric or polarisation tensor) in $\mathcal{K}'$ is assumed to be the sum of the contributions found by Lorentz transforming the response tensor for each species from its rest frame to the pulsar frame.

\subsection{Inclusion of the pulsar current}

With this relabeling the primes on $z',y',\beta'_\pm, \theta'$ are retained, as in the original notation. However, unprimed parameters in the original rest frame $\mathcal{K}$ are now different in each of the two frames $\tr{\mathcal{K}}{\epsilon}$. Relevant parameters in the two frames are denoted by adding a superscript $\epsilon$ to indicate the frame. Specifically, we make the replacements $z,y,\beta_\alpha,\theta\to \tR{z}{\epsilon},\tR{y}{\epsilon},\tRr{\beta}{\epsilon}{\alpha},\tR{\theta}{\epsilon}$.

The Lorentz transformations between the rest frames and the pulsar frame imply
\begin{equation}
	\tR{z}{\epsilon}=\frac{z'-\tRr{\beta}{\epsilon}{\rm s}}{1-z'\tRr{\beta}{\epsilon}{\rm s}},
	\qquad
	\tRr{\beta}{\epsilon}{\pm}=\frac{\beta'_\pm-\tRr{\beta}{\epsilon}{\rm s}}{1-\beta'_\pm\tRr{\beta}{\epsilon}{\rm s}},
	\qquad
	\tR{y}{\epsilon}=\frac{y'}{\tRr{\gamma}{\epsilon}{\rm s}(1-z'\tRr{\beta}{\epsilon}{\rm s})},
	\qquad
	\tan\tR{\theta}{\epsilon}=\frac{\tan\theta'}{\tRr{\gamma}{\epsilon}{\rm s}(1-z'\tRr{\beta}{\epsilon}{\rm s})}.
	\label{betazpm}
\end{equation}

\subsection{Response tensor including the pulsar current}

The response tensor in the form \eqref{Piij1a} involves functions and RPDFs defined in the original rest frame $\mathcal{K}$, in the case where both distributions have the same streaming speed. The generalization to the case where the two rest frames $\tr{\mathcal{K}}{\epsilon}$ are different follows by adding a subscript $\epsilon$ to the relevant parameters in each frame. The response tensor in this case generalizes to
\begin{align}
	 & {\Pi'}_{11} ={\Pi'}_{22}
	=-\sum_{\epsilon}\frac{e^2\tR{n}{\epsilon}}{m}\frac{1}{1+\tR{y} {\epsilon 2}}\left[\tr{\aV{\frac{1}{\gamma}}}{\epsilon}+\sum_{\alpha=\pm}\frac{\alpha(\tR{z}{\epsilon}-\tRr{\beta}{\epsilon}{\alpha})^2\tr{R}{\epsilon}(\tRr{\beta}{\epsilon}{\alpha})}{\tRr{\beta}{\epsilon}{+}-\tRr{\beta}{\epsilon}{-}}\right],
	\nn                                                                                                                                                                                                                                                                                                         \\
	 & {\Pi'}_{33}=-\tRr{\gamma }{\epsilon 2}{\rm s}\sum_{\epsilon}\frac{e^2\tR{n}{\epsilon}}{m}\left\{(\tR{z}{\epsilon}+\tRr{\beta}{\epsilon}{\rm s})^2\tr{W}{\epsilon}(\tR{z}{\epsilon})+\frac{\tan^2\tR{\theta}{\epsilon}}{1+\tR{y}{\epsilon 2}}\left[\tr{\aV{\frac{1}{\gamma}}}{\epsilon}+\sum_{\alpha=\pm}\frac{\alpha(\tRr{\beta}{\epsilon}{\rm s}+\tRr{\beta}{\epsilon}{\alpha})^2\tr{R}{\epsilon}(\tRr{\beta}{\epsilon}{\alpha})}{\tRr{\beta}{\epsilon}{+}-\tRr{\beta}{\epsilon}{-}}\right]
	\right\},
	\nn                                                                                                                                                                                                                                                                                                         \\
	 & {\Pi'}_{13} ={\Pi'}_{31}
	=-\tRr{\gamma}{\epsilon}{\rm s}\sum_{\epsilon}\frac{e^2\tR{n}{\epsilon}}{m}\frac{\tan\tR{\theta}{\epsilon}}{1+\tR{y}{\epsilon 2}}\left[-\tr{\aV{\frac{1}{\gamma}}}{\epsilon}
		+\sum_{\alpha=\pm}\frac{\alpha(\tR{z}{\epsilon}-\tRr{\beta}{\epsilon}{\alpha})(\tRr{\beta}{\epsilon}{\rm s}+\tRr{\beta}{\epsilon}{\alpha})\tr{R}{\epsilon}(\tRr{\beta}{\epsilon}{\alpha})}{\tRr{\beta}{\epsilon}{+}-\tRr{\beta}{\epsilon}{-}}
		\right],
	\nn                                                                                                                                                                                                                                                                                                         \\
	 & {\Pi'}_{12} =-{\Pi'}_{21}
	=-\rmi\sum_{\epsilon}\frac{\epsilon e^2\tR{n}{\epsilon}}{m}\frac{\tR{y}{\epsilon}}{1+\tR{y}{\epsilon 2}}\sum_{\alpha=\pm}\frac{\alpha(\tR{z}{\epsilon}-\tRr{\beta}{\epsilon}{\alpha})\tr{S}{\epsilon}(\tRr{\beta}{\epsilon}{\alpha})}{\tRr{\beta}{\epsilon}{+}-\tRr{\beta}{\epsilon}{-}},
	\nn                                                                                                                                                                                                                                                                                                         \\
	 & {\Pi'}_{23} =-{\Pi'}_{32}
	=\rmi\tRr{\gamma}{\epsilon}{\rm s}\sum_{\epsilon}\frac{\epsilon e^2\tR{n}{\epsilon}}{m}\frac{\tR{y}{\epsilon} \tan\tR{\theta}{\epsilon}}{1+\tR{y}{\epsilon 2}}\sum_{\alpha=\pm}\frac{\alpha(\tRr{\beta}{\epsilon}{\rm s}+\tRr{\beta}{\epsilon}{\alpha})\tr{S}{\epsilon}(\tRr{\beta}{\epsilon}{\alpha})}{\tRr{\beta}{\epsilon}{+}-\tRr{\beta}{\epsilon}{-}}.
	\label{Piij1aJ}
\end{align}
Using the relations \eqref{betazpm} the response tensor \eqref{Piij1aJ} may be rewritten in terms of the parameters $ \tR{z}{\epsilon \prime}, \tR{y}{\epsilon \prime}, \tRr{\beta}{\epsilon \prime}{\alpha}, \tR{\theta}{\epsilon \prime}$ in the pulsar frames $\mathcal{K}'$.

\end{document}